\documentclass[pre,showpacs,notitlepage,twocolumn]{revtex4-1}
\pdfoutput=1

\usepackage[colorlinks,linkcolor={blue},citecolor={blue},urlcolor={blue}]{hyperref}

\usepackage{graphicx}

\usepackage{bm,amsmath}

\usepackage{latexsym}

\usepackage{verbatim}

\usepackage{color}

\usepackage{subfigure}

\usepackage{gensymb}

\usepackage{soul}

\usepackage[dvipsnames]{xcolor}

\newcommand{\bR}{\mathbf{R}}

\begin{document}


\title[]{ Static and dynamic scaling behavior of a  polymer melt model with triple-well bending potential }

\author{Sara Jabbari-Farouji $^{*}$}
\affiliation{Institute of Physics, Johannes Gutenberg-University, Staudingerweg 7-9,
55128 Mainz, Germany
}%
 \email{Correspondence to: sjabbari@uni-mainz.de}


 \date{\today}


  \begin{abstract}
  We perform molecular-dynamics simulations for  polymer melts of a coarse-grained polyvinyl alcohol model that crystallizes upon slow cooling. To establish the properties of its   high temperature liquid state as a reference point, we characterize in detail the structural  features of   equilibrated polymer melts with chain lengths $5\le N \le 1000$ at a temperature slightly above their crystallization temperature. We find that the conformations of sufficiently long polymers with $N >50$ obey essentially the  Flory's ideality hypothesis. The chain length dependence of the end-to-end distance and the gyration radius follow   the scaling predictions of ideal chains and  the probability distributions of the end-to-end distance, and  form factors are in good agreement with  those of ideal chains. The intrachain correlations reveal  evidences for incomplete screening of self-interactions. However, the observed deviations are small. Our results rule out any pre-ordering or mesophase  structure formation that are proposed as precursors of polymer crystallization in the melt. Moreover, we   characterize in detail  primitive paths of long  entangled   polymer melts  and we examine scaling predictions of Rouse and the reptation theory for the mean squared displacement of  monomers and polymers center of mass.
   

\end{abstract}


\maketitle

\section* {Introduction}

Polymer melts  are dense liquids  consisting merely of macromolecular chains. The main characteristic of polymer melts is their high-packing density
which leads to  overlapping of pervaded volume of their chains \cite{Rubinstein}. As a result, density fluctuations in a melt are small and  similar to a simple fluid every monomer is isotropically surrounded by other monomers that can be part
of the same chain or belong to other chains. As a consequence the chains conformations in the molten state are close to random walks.
Nevertheless, the crystallization of polymer melts   remarkably differs from monomeric liquids due  to chain connectivity constraints that must be compatible with lattice spacings of
a crystalline structure. 
Upon  cooling of a crystallizable polymer melt, a semicrystalline structure emerges that comprises  of  regularly packed, extended chain 
sections surrounded by amorphous strands \cite{semicrys}. 

Despite intensive research, the mechanism of polymer crystallization is still poorly understood as 
this process is determined by an intricate interplay of  kinetic effects  and thermodynamic driving force 
\cite{Keller,Sommercrys}. From theoretical perspective, attempts have been made to understand thermodynamics of polymer crystallization. 
These efforts include lattice-based model \cite{Florycrys}, Landau-de Gennes  types of approach \cite{Olmsted} and 
density functional theory \cite{DFTPaul,DFT1}. Among these, the density-functional theory 
is a particularly promising approach as it  can predict equilibrium features of  inhomogeneous   semicrystalline structure from the 
homogeneous melt  structure  provided  that a suitable free-energy functional is employed and accurate information about chain conformations and structure of polymeric liquids
is supplied \cite{DFTreview}.  Such an input can be obtained from simulations of a crystallizable polymer melt that allow us to compute the intrachain and interchain  correlation functions. Here, our aim is to characterize conformations of a crystallizable polymer melt model known as  coarse-grained polyvinyl alcohol (CG-PVA) \cite{Meyer2001}. Owing to its coarse-grained nature, this model allows for simulations of large-scale structure of semicrystalline polymers.
  
  The  CG-PVA  is a  bead-spring polymer model  that  is obtained by a systematic coarse-graining from atomistic simulations of
  polyvinyl alcohol \cite{Reith}.  The coarse-graining procedure maps all the atoms of a monomer into one bead and a bending potential is extracted from the bond-angle distribution of atomistic simulations of PVA polymers.  The main distinctive feature of the model is its triple-well  intrachain bending rigidity  that leads to formation of chain-folded crystallites upon slow cooling of the melt \cite{Meyer2001,SaraMacro}. Although the model does not take into account the intermolecular hydrogen bonding interactions, it reproduces the main features of polymer crystallization  such as lamella formation.  Recently, the CG-PVA  model has been employed to study the dependence of polymer crystallization on the chain length \cite{SaraPRL2017,Joerg2016}.
  To understand the influence  of chain length  and entanglements on the  melt crystallization \cite{Luo2013,Luo2016}, the  properties of its high-temperature liquid state need to be firmly established as a reference point. Here, we focus on the static and dynamic properties of CG-PVA polymer melts  at a temperature which is 1.1 times the crystallization temperature.

  The prior studies of structural properties  of  this system  have been limited to static properties of short chains  $N \le 100$ \cite{Vettorel2007}  with purely repulsive interactions.  Here, we  present the results for CG-PVA  polymer melts with $5 \le N \le 1000$  with attractive interactions.  
  Characterizing the  conformational features of longer polymer melts, we  accurately determine the persistence length, the Flory's characteristic ratio and  the entanglement  length of   CG-PVA polymers. Examining carefully the consequences of  the triple-well  bending potential on  chain  conformations, we   barely find any evidence of pre-ordering or microstructure formation as precursors of crystallization. 
  Encompassing  a crossover from short unentangled chains to  the entangled ones,  we examine  the   credibility of Flory's ideality hypothesis and manifestations of incomplete screening of excluded volume and hydrodynamic interactions on the structural and dynamical features of entangled polymer melts with a finite persistence length.

 Flory's ideality  hypothesis  states that polymer conformations in a melt behave statistically as ideal random-walks on length scales much larger than the monomer's diameter \cite{Flory,polymerDoi}.
 This ideality hypothesis is a mean-field result that relies on the negligibility of density fluctuations in polymer melts.
Therefore, its validity  is  not taken for granted. Indeed, the computational studies of fully flexible long polymers for both lattice (bond fluctuation) and continuum  (bead-spring)  models  have revealed  noticeable deviations from the ideal chain behavior   \cite{Wittmer2004,Wittmer2007a,Wittmer2007b,formfactor2007,Hsu2014}. 
The theoretical calculations show that these  deviations result from the interplay between the chain connectivity and the melt incompressibility  which foster an incomplete screening of excluded volume 
 interactions  \cite{Wittmer2007a,formfactor2007,Semenov}.  However, a recent study of conformational properties of long locally semiflexible polymer melts demonstrates that the deviations
 diminish  as the chains bending stiffness  increases and the conformations of sufficiently stiff  chains  are well described by  the theoretical predictions for
 ideal chains \cite{Kremer2016}.

 Investigation of the static scaling behavior  of locally semiflexible CG-PVA  polymer melts confirms that  CG-PVA polymers display globally random-walk like conformations.
 Notably, for chains with $N>50$, the results of the end-to-end distance, gyration radius, the probability distribution functions  and the chain structure factor are in good agreement with 
the theoretical predictions for ideal chains. However, inspection of intrachain correlations reveals some evidences for deviations from ideality. The mean-square internal distances of long chains $N>100$  are slightly swollen compared to ideal chains due to incomplete
screening of excluded volume interactions. Additionally, the second Legendre polynomial of angle between bond vectors exhibits a power law decay for   curvilinear distances  larger than the persistence length providing another
testimony for self-interaction of chains. Nonetheless, we note that these visible deviations are small.

 The remainder of the paper is organized as follows. In Sec. II, we briefly review  the CG-PVA model and 
provide the simulation details. We present  a detailed analysis of  conformational and structural features of polymer melts  in Sec. III and we compare simulation results to the theoretical predictions for ideal chains.
We investigate  conformational properties of the primitive paths of
long chains  in section IV where we  determine the  entanglement length of fully equilibrated CG-PVA chains.
Sec. V explores the segmental motion of polymers  at different characteristic timescales and examines the scaling
laws predicted by the Rouse model and the reptation theory \cite{Rubinstein,polymerDeGennes,polymerDoi}.
Finally,  we summarize our main findings and discuss our future directions  in section VI.

\section*{ Model and simulation details}
We equilibrate  polymer melt configurations of the  coarse-grained polyvinyl alcohol (CG-PVA)  model using molecular dynamics simulations. In the following, we first briefly review CG-PVA model and then provide the details of  simulations.
 \subsection*{Recap of  the CG-PVA model}
In the CG-PVA  bead-spring model, each  bead of the coarse-grained chain
with diameter $\sigma=0.52$ nm corresponds to  a monomer of the PVA polymer. The fluctuations of the bond length about  its average value  $b_0=0.5 \sigma$  are restricted by a  harmonic potential with a bond stiffness  constant $k_{bond}=2700 k_B T/\sigma^2$
\begin{equation}
U_{bond}=\frac{1}{2} k_{bond}(b-b_0)^2
 \end{equation}
 that leads to bond length fluctuations  with a size much smaller than the monomer diameter. 
Monomers of distinct chains and the same chain  that are three bonds or farther apart interact by  a soft 6-9 Lennard-Jones  potential,
\begin{equation}
U_{nb}(r)= \epsilon_0 \left[  (\frac{\sigma_0}{r})^9 -(\frac{\sigma_0}{r})^6\right]
 \end{equation}
in which $\sigma_0=0.89 \sigma$ and $\epsilon_{0}=1.511$ $k_BT_0$ . Here,   $T_0=550$ K is the reference temperature of the PVA melt \cite{Meyer2001}. We truncate  and shift the Lennard-Jones potential at $r^C=1.6 \sigma$ in our simulations.  Our choice of $r^C$ is different from initial studies where the non-bonded interactions were truncated at the minimum of the LJ potential $r_{min}\approx1.02 \sigma$ and thus were purely repulsive. The purely repulsive model was initially used to study structure formation in the quiescent state \cite{Meyer2001,Meyer2002,Luo2013,Luo2016}.
The attractive part is needed for non-equilibrium studies of deformation \cite{SaraMacro,SaraLMC,SaraPRL2017}.
Note that the choice of $r^C=1.6$ leads to an increase of density by about 10 \%  with respect to the purely repulsive conditions.
However, the structural properties of the melt remain essentially unaffected.

\begin{figure}[t]
\includegraphics[width=0.98\linewidth]{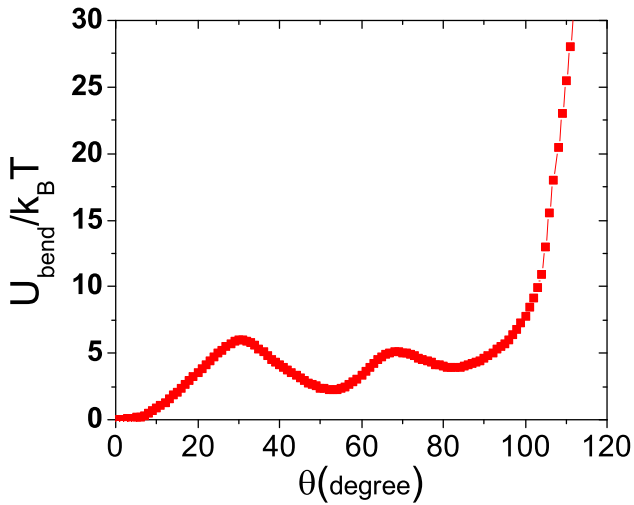}
\caption{The angular dependence of bending potential  $U_{bend} (\theta)$ for CG-PVA model as a function of the angle $\theta$ between two successive bond vectors. }
\label{fig1} 
\end{figure}

 The distinguishing characteristic of the CG-PVA model  is its triple-well angle-bending potential   \cite{Meyer2001} as presented in Fig. \ref{fig1}. 
 This  bond angle potential is determined directly from atomistic simulations  by Boltzmann inversion of  the probability distribution of the bond angle  $\theta$ \cite{Reith,Meyer2001}. The  minima of   $U_{bend}(\theta)$ reflect the specific states of two successive torsion angles at the atomistic level and they correspond to three energetically favorable states, {\it trans-trans, trans-gauche and gauche-gauche}. 
 Therefore, the bending potential  retains semiflexibility of chains   originated from the torsional states of the   atomistic backbone \cite {Meyer2001}. 


  %

\subsection*{Units and simulation aspects}
\begin{table}[t]
\begin{tabular}{ | c | c | c | c | c| c | c| }
 \hline
$N$ & $n_c$ & $t_{\text{eq}}/\tau \times 10^6$ & $\rho_{m} \sigma^3 $ &$R_g/\sigma$ & $R_e/\sigma$ \\
  \hline
  $5$ & $1000$  & $0.05$ & 2.06 &  0.615 & 1.63  \\
   \hline
 $10$ & $1000$  & $0.05$ & 2.20 & 1.08  & 2.98 \\
 \hline
 $20$ & $1000$  & $0.10$ & 2.28 &  1.79 & 4.79\\
 \hline
 $30$ & $1000$ & $0.15$ & 2.30 &  2.35 & 6.09\\
 \hline
 $50$ & $1000$  & $0.32$  &  2.33 & 3.17  & 8.11\\
 \hline
 $100$ & $320$  & $0.9$ & 2.34 & 4.69  &  11.73\\
 \hline
 $200$ & $500$  & $2.5$ & 2.35 & 6.68  & 16.53\\ 
 \hline
 $300$ & $300$  & $6$ & $2.35$ & 8.25   & 20.31 \\ 
 \hline
 $500$ & $500$  & $20$ &  2.35 & 10.85  & 26.49\\  
 \hline
 $1000$ & $500$  & $15$  &  2.36 & 15.11  & 37.27\\ 
   \hline
   \end{tabular}
   \caption{Summary of polymer melts configurations  produced in the simulations and the equilibration time before the measurements were started   $t_{\text{eq}}$.  $N$ is the number of beads in
   a chain,  $n_c$ is the number of chains. We also report the values of  average monomer density $\rho_m$, average  gyration radius $R_g=\sqrt{\langle R_g^2 \rangle }$ and end-to-end distance $R_e=\sqrt{\langle R_e^2 \rangle }$ of the equilibrated polymer conformations. }
\end{table} 

We carry out  molecular dynamics simulations of CG-PVA polymers  with chains lengths $5 \le N \le 1000$ using LAMMPS~\cite{LAMMPS}.   We report the distances in length unit $\sigma=0.52$ nm.  \cite{Meyer2001}.  The time unit from 
the conversion relation  of units   is $\tau=\sqrt{m \sigma^2/k_B T} $ with the monomer mass $m=1$.
The starting melt configurations are  prepared by generating an ensemble of  $n_c$ (number of chains) self-avoiding random walks composed of  $N$ monomers  with an initial density of $\rho \sigma^3=2.31$.  To remove 
the interchain monomer overlaps, we use a slow push off method in which initially  a very weak LJ potential with $\sigma=0.5 \sigma_0$ and a very small $\epsilon_{LJ}=10^{-3}$ is switched on. Next, the range and strength of Lennard-Jones potential are gradually  increased  to their final values.   
We equilibrate  disordered melt structures in the NPT ensemble using a Berendsen barostat  and  a Langevin thermostat
with friction constant $\Gamma=0.5$. The temperatures and pressures are reported in  reduced units  $T=1.0$ and $P=8$,  equivalent to $T_0=550$ K  and  $P_0=1$ bar in atomistic simulations. The time step  used through all the simulations is $0.005 \tau$. The chosen melt temperature is about 10\% higher than the crystallization temperature $T_c\approx 0.9$ of long   polymer melts  ($N \ge 50$) obtained at a cooling-rate of $10^{-6}$ $\tau^{-1}$ \cite{SaraPRL2017}.

The polymer configurations for $N \le 500$ were equilibrated until the average  monomers mean-square  displacement
$\langle \Delta r^2(t) \rangle$ is equal  or larger than their mean square end-to-end distance $\langle R_e^2 \rangle$. The time for which $\langle \Delta r^2(t) \rangle=\langle R_e^2 \rangle$ is a measure of the relaxation 
time   and it is comparable to the Rouse 
time for the short chains and the disengagement time for the entangled chains \cite{polymerDoi} as will be verified in section V. Polymer melts with $N=1000$,  were equilibrated until $\langle \Delta r^2(t) \rangle$
is comparable to their mean-square gyration radius $\langle R_g^2 \rangle$.  Table 1 provides a summary of configurations of polymers and the simulation time for equilibration  after push off stage  in units of $\tau$. Then, the  data for characterization
of   static  and dynamic properties are acquired.

 In order to analyze the  static properties  of CG-PVA  polymers,  we  extract from the polymer  configurations  the normalized probability distribution functions of the internal distances, the gyration radius, the bond  length and angle of 
  polymers as well  as the bond  length and angle of their primitive paths.
We  acquire the numerical  probability distribution  of a desired observable $x$  by accumulating a histogram  $H_N(x)$ of a fixed width $\Delta x$. Then, we calculate the normalized  probability distribution function as $P_N(x)=\frac{H_N(x)}{\sum_{x'} H_N(x') \Delta x}$. 

 %
\begin{figure}[t]
\includegraphics[width=0.98\linewidth]{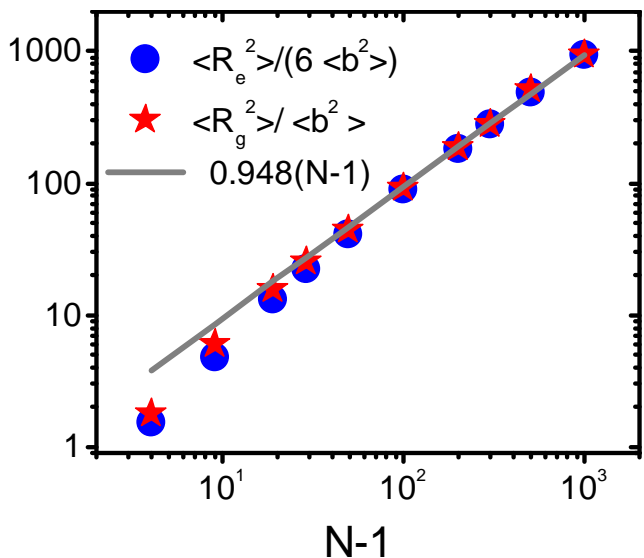}
\caption{Log-log plot of rescaled   mean square end-to-end distance $\langle R_e^2 \rangle /(6 \langle b^2 \rangle)$ and gyration radius $\langle R_g^2\rangle/ \langle b^2 \rangle $  versus $N$ where $\langle b^2 \rangle=0.247$. 
The solid line shows a linear fit of $\langle R_e^2\rangle/ 6 \langle  b^2 \rangle $ for $N > 50$. Hence, sufficiently long CG-PVA polymers in a melt  follow  $\langle R_e^2\rangle \propto \langle R_g^2\rangle\approx  N^{2\nu}$  
 with the scaling  exponent $\nu=1/2$ in agreement to that of ideal chains.
}
\label{fig2} 
\end{figure}
\begin{figure*}[ht]
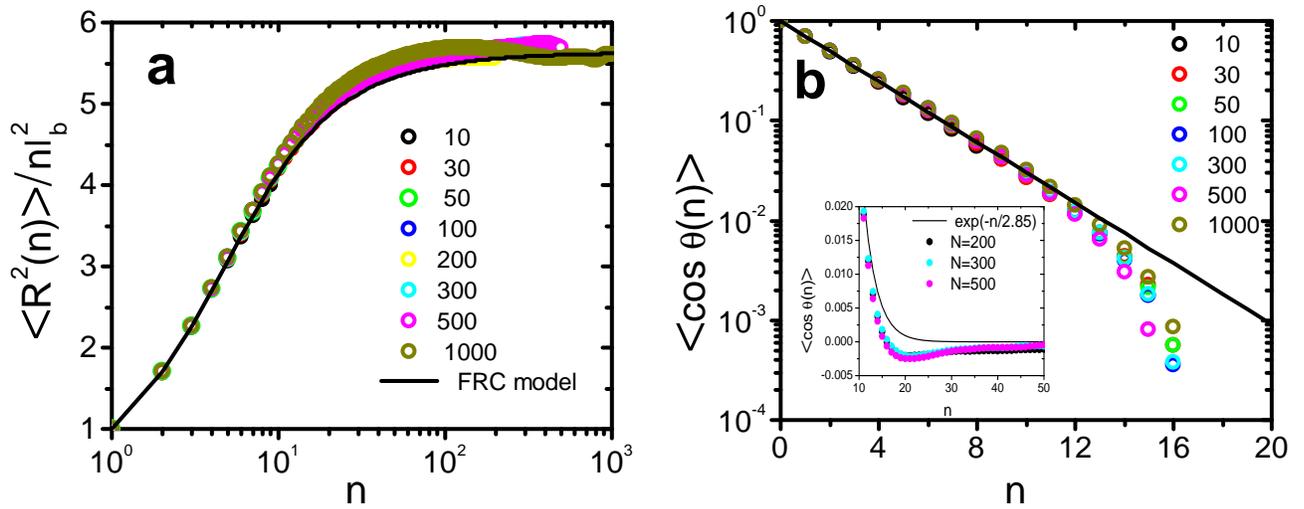

\includegraphics[width=0.48\linewidth]{fig3a.pdf}
\includegraphics[width=0.48\linewidth]{fig3b.pdf}
\caption{(a) Rescaled mean square internal distance, $\langle R^2(n) \rangle/ n \ell_{b}^2  $  plotted as a
function of the  curvilinear distance  $n$ along the chain backbones for several chain lengths. The solid line shows the theoretical prediction of  the generalized freely rotating
chain (FRC) model with $ \langle \cos \theta \rangle=0.699 $. (b)  Semilog plot of bond-bond orientational correlation function $ \langle \cos \theta  (n) \rangle$ versus $n$ for various chain sizes.
The straight line shows the fit with exponential decay of the form $\exp(-\ell_b/\ell_p n$) with $\ell_b/\ell_p=0.35 \pm 0.01$.  The inset shows a linear-linear  plot of diminishing bond-bond orientational correlation function of
$N=200$, 300 and 500 for $ 10 < n \le 40$.}
\label{fig3} 
\end{figure*}

\section*{Structural features of CG-PVA polymer melts}

We first present the results on the chain-length dependence of the mean square end-to-end distance and the gyration radius for chain sizes in the range $ 5\le N \le 1000$. 
The mean square end-to-end distance is defined as
\begin{equation}
 \langle R_e^2 \rangle=\frac{1}{n_c}  \sum_{i=1}^{n_c} \langle(\mathbf{r}_{i,N}-\mathbf{r}_{i,1})^2 \rangle
\end{equation}
and the mean square gyration radius  is given by
\begin{equation}
 \langle R_g^2 \rangle=\frac{1}{n_c N}  \sum_{i=1}^{n_c} \langle \sum_{n=1}^{N} (\mathbf{r}_{i,n}-\mathbf{r}_{i,cm})^2 \rangle
\end{equation}
where $\mathbf{r}_{i,n}$ is the position of  $n$th monomer of chain number $i$ and $\mathbf{r}_{i,cm}$ is the center of mass position  of the $i$th polymer chain in a sample. Here,   $\langle \cdots \rangle$  
includes an averaging over $50-200$ equilibrated configurations that are     $1-5 \times 10^3\tau$  apart for the shorter chains and  $5 \times 10^4\tau$ apart for the longer chains with $N>200$. 
Fig. \ref{fig2} shows $\langle R_g^2\rangle/ \langle \mathbf{b}^2 \rangle $ and  $\langle R_e^2\rangle/(6 \langle \mathbf{b}^2 \rangle) $
as a function of chain length $N$.  Here, $\langle \mathbf{b}^2 \rangle:=\ell_b^2=0.2477 \pm 0.0002$  denotes the mean-square bond length  that is  independent of  the chain length.
The longer chains with $N>50$ follow the  relation $\langle R_e^2\rangle/\langle R_g^2\rangle=6$  valid for ideal chains \cite{polymerDeGennes}. For shorter chains the ratio $\langle R_e^2\rangle/\langle R_g^2\rangle$ is in the range $6.5-7.5$.
Additionally, chains with $N > 50$ follow the scaling behavior of ideal chains $\langle R_e^2\rangle \propto \langle R_g^2\rangle \propto  N$. The extracted  scaling exponent from fitting  $\langle R_e^2\rangle \propto N^{2 \nu}$
versus $N$ with a power law  gives $\nu=0.50 $ that is  identical to the value for the ideal chains.
The observed scaling behavior for the  mean square end-to-end distance and the gyration radius  of CG-PVA polymers with $N>50$ suggests that long    polymers behave like ideal chains. 
 In the following, we investigate in more detail the  conformational statistics of individual chains, and compare them to the theoretical predictions for ideal chains \cite{polymerDeGennes,Rubinstein} .

\subsection*{ Intrachain correlations}
\begin{figure}[t]
 \includegraphics[width=0.98\linewidth]{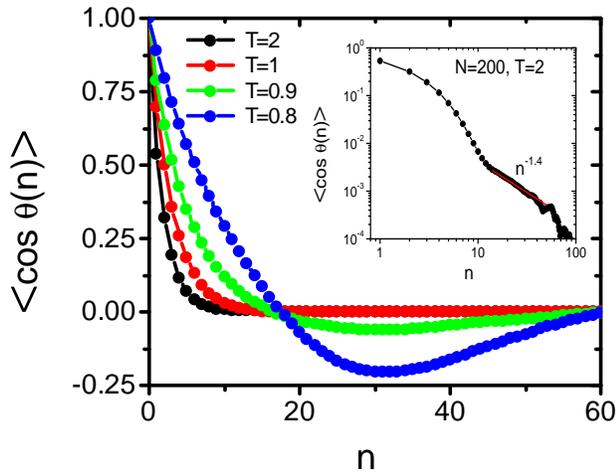}
\caption{Bond-bond orientational correlation function $ \langle \cos \theta  (n) \rangle$ of CG-PVA polymers of chain length $N=200$ versus $n$ at different temperatures as given in the legend. The samples at $T=0.9$ and 0.8 are obtained from cooling of melt at $T=1$  with constant rate of $10^{-6}$ $\tau$. The inset depicts a log-log plot of $ \langle \cos \theta  (n) \rangle$ at $T=2$ for which a power law decay at $n>10$ emerges.}
\label{fig4} 
\end{figure}
We begin by characterizing the  intrachain correlations for both monomer positions and bond orientations. To quantify the positional intrachain correlations,  we calculate the mean-square internal distances (MSID) for various chain lengths  defined as 
\begin{equation}
 \langle R^2(n) \rangle=\frac{1}{n_c} \left\langle \sum_{i=1}^{n_c}  \sum_{j=1}^{N-n} \frac{1}{N-n}(\mathbf{r}_{i,j}-\mathbf{r}_{i,j+n})^2 \right\rangle
\end{equation}
where $n$ is the curvilinear (chemical) distance between the $j$th monomer and
the $(j+n)$th monomer along the same chain.  MSID is a  measure of
internal chain conformation that can be used to evaluate the equilibration degree of long polymers.

In Fig. \ref{fig3}a, we present  the rescaled mean square internal distance, $\langle R^2(n) \rangle/ n  \ell_b^2 $  obtained by averaging over  $50-200$ polymer melt configurations
that are  $10^3$ $\tau$ ($5\times10^4$ $\tau$) apart for short (long) chains. Up to $n \approx 10$, we find a good collapse of all the MSIDs. However, for larger curvilinear distances,
the MSIDS of  longer chains  $N >100$ are slightly larger but   all follow the same master curves  suggesting that longer chains are a bit swollen due to chains self-interactions.
Note that the deviations of $\langle R^2(n) \rangle/n \ell_b^2$ of   $N=1000$ chains  for curvilinear distances $n>100$ from other long chains are due to their poor equilibration.
These results show that the $N=1000$ chains are   equilibrated at shorter length scales but their large-scale conformation still needs much longer equilibration time  $ \tau_{eq} \propto 3 N^3 /N_e\tau \approx 10^8 \tau$ where $N_e \approx 36$ is
the entanglement length as will be discussed in the  section IV. 

From the asymptotic behavior of mean square end-to-end distances of long CG-PVA chains, we can extract their  characteristic ratio $C_{\infty}$ and  Kuhn  length $\ell_K$.   The characteristic ratio is
defined by the relation   $\langle R_e^2(N) \rangle= C_{\infty} (N-1) \ell_b^2$ where $\ell_b$ is the average bond length. From MSID of longer polymers, $N=300$ and $N=500$, we 
obtain $ C_{\infty} =5.70 \pm 0.04$.  The Kuhn length gives us the effective bond length of an equivalent freely jointed chain which has the same 
 mean square end-to-end distance $R_e^2$  and the same maximum end-to-end to distance $R_{\text{max}}$ \cite{Rubinstein}. For a freely jointed chain with $N_k$ Kuhn segments with  bond length $\ell_K$, we have
 $R_{\text{max}}=(N_k-1) \ell_k$ and $\langle R_e^2(N_k) \rangle=  (N_k-1)  \ell_k^2$. For  CG-PVA polymers, we find  $R_{\text{max}}=(N-1) \ell_b$ and  $\langle R_e^2(N \gg 1) \rangle= 5.70 (N-1)  \ell_b^2$. 
  Equating $\langle R_e^2(N) \rangle$ and $R_{\text{max}}$  of the CG-PVA chains with those of the equivalent  freely jointed chain, we obtain $\ell_k=5.70 \ell_b=2.83$  $\sigma$.


Next, we compare  the mean square internal distance of CG-PVA  polymers with $\langle R^2(n) \rangle$ of the generalized freely rotating chain (FRC) model \cite{Flory,FRC} for ideal chains.
If the  excluded volume interactions between different parts of a certain polymer are screened, one expects   the FRC  model to provide  a good description of CG-PVA polymer melts. 
$\langle R^2(n) \rangle$ of  FRC  model only depends on the   value of $ \langle \cos \theta \rangle$ where  $\theta$ is   the  angle between any two successive bonds in
a chain. It is given by 
\begin{equation}
 \langle R^2(n) \rangle=n\ell_b^2 \left( \frac{1+\langle \cos \theta \rangle}{1-\langle \cos \theta \rangle}-\frac{2}{n} \frac{\langle \cos \theta \rangle (1-\langle \cos \theta \rangle^n)}{(1-\langle \cos \theta \rangle)^2}   
 \      \right) .
 \label{eq:Boltzmann}
\end{equation}

The   value of  $ \langle \cos \theta \rangle $ for the CG-PVA model  can be obtained from $P_N(\theta) \propto \exp[-\beta U_{bend} (\theta)] $  as
\begin{equation}
 \langle \cos \theta \rangle=   \frac{\int_{0}^{\pi} \sin \theta \cos \theta \exp[-\beta U_{bend} (\theta)] }{\int_{0}^{\pi} \sin \theta  \exp[-\beta U_{bend} (\theta)]}.        
\end{equation}
where $U_{bend} (\theta)$ is presented in Fig. \ref{fig1}.
Doing the  integration  in Eq. \eqref{eq:Boltzmann} at $T=1.0$ numerically, we obtain 
 $  \langle \cos \theta \rangle= 0.6985$.    We can also  directly  infer $ \langle \cos \theta \rangle $ from  the MD simulations results as $ \langle \cos \theta \rangle \equiv \langle \widehat{\mathbf{b}}_{i,j} \cdot \widehat{\mathbf{b}}_{i,j+1}  \rangle $ where  $\widehat{\mathbf{b}}_{i,j}$  is  the  $j$th unit bond vector   of the  $i$th chain and the averaging is carried out over all the chains and 100-500 equilibrated polymer melt configurations that are $10^3$ $\tau$ apart.
From MD simulations, we deduce the universal value of $ \langle \cos \theta \rangle=0.699 \pm 0.005 $ independent of the chain length.
This value  agrees well with the Boltzmann-averaged mean value.

  \begin{figure*}[t]
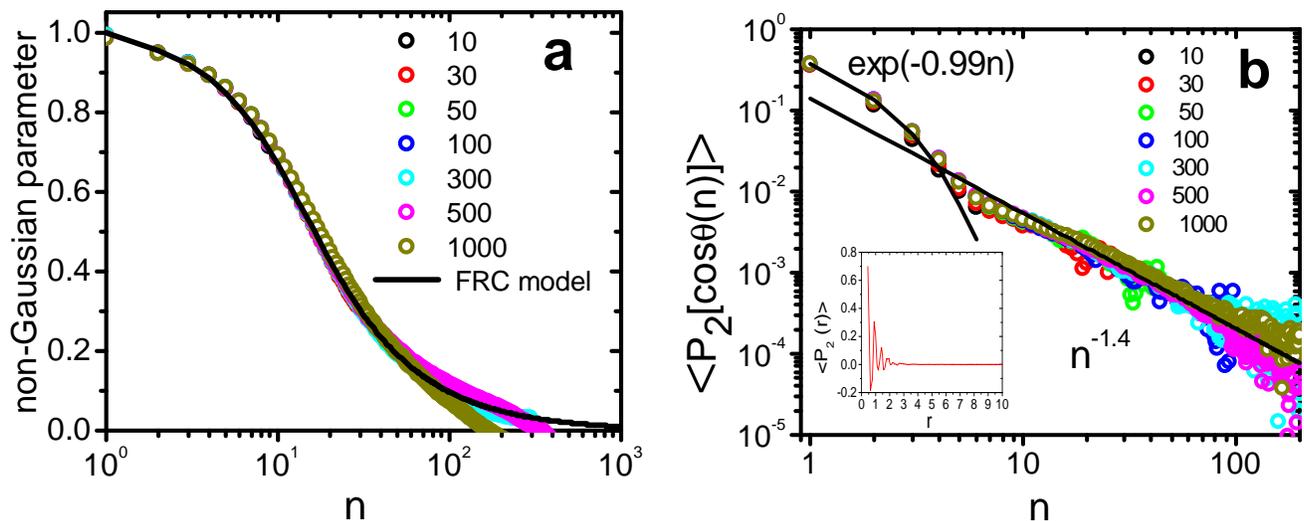

\includegraphics[width=0.48\linewidth]{fig5a.pdf}
\includegraphics[width=0.48\linewidth]{fig5b.pdf}
 \caption{  (a) Non-Gaussian parameter defined as $\alpha(n)=1/2(5-3\langle R^4(n) \rangle /  \langle R^2(n) \rangle ^2 ) $  plotted as a
function of the  curvilinear distance $n$ for several chain lengths. The solid line shows the theoretical prediction of  the generalized freely rotating
chain (FRC) model with $ \langle \cos \theta \rangle=0.699 $ and $\langle P_2[ \cos \theta]\rangle=0.37$. (b) Average of the second Legendre polynomial  of cosine of
angle between any two bonds with a curvilinear distance $n$, $\langle P_2[ \cos \theta(n)]\rangle $,   decays exponentially versus $n$ for small  curvilinear distances $n<5$, and it exhibits a power law for larger $n$ values.  
The inset shows that the average of second Legendre polynomial of cosine of angle between two bonds  with a separation   $r$  oscillates strongly and decays rapidly with distance.
   }
\label{fig5} 
\end{figure*}

In Fig. \ref{fig3}a, we have also included the MSID of an equivalent freely rotating chain with $ \langle \cos \theta \rangle=0.699$.  We find that   MSIDs of short chains $N\le 100$ 
fully agree with that of the  freely rotating chain  model  whereas   the MSIDs of longer chains present noticeable  deviations from the FRC theory for $n >10 $.
Hence, FRC model slightly underestimates the  MSID of longer chains. Likewise, the characteristic ratio of  the FRC model, given by
$C_{\infty}=\frac{1+\langle \cos \theta \rangle}{1-\langle \cos \theta \rangle}=5.64$   is slightly lower than the $C_{\infty}=5.70$ estimated from the simulation results.   
These very small deviations are most-likely due to the correlation hole effect that stems from
incomplete screening of  interchain excluded volume interactions and leads to long-range intrachain correlations \cite{polymerDeGennes}. 
 
 The observed  swelling of long chains suggests that  an evidence of remnant long-range  bond-bond correlations should be detectable. Thus, we  examine the intrachain  orientational bond-bond correlations  
$ \langle \cos \theta  (n) \rangle \equiv \langle \widehat{\mathbf{b}}_{i,j} \cdot \widehat{\mathbf{b}}_{i,j+n}  \rangle $  as a 
 function of the internal distance $1 \le n \le N-1$.  Fig. \ref{fig3}b represents the orientational bond-bond correlations for different chain lengths that are obtained by averaging over 500-2500 equilibrated configurations that are $10^3 \tau$ apart.  
 We find that the   bond-bond correlation functions of all the chain lengths decay exponentially up to $n \approx  10$  and  can be
well described by $\exp(-0.35 n)$.  We   extract the so-called persistence length $\ell_p$ from the bond-bond orientational correlation functions.  $\ell_p$ 
is defined by their decay length, more precisely:
\begin{equation}
  \langle \widehat{\mathbf{b}}_{i,j} \cdot \widehat{\mathbf{b}}_{i,j+n}  \rangle= \exp(-n \ell_b/\ell_p).
  \end{equation}
 Using $\ell_b=0.497$ and $\ell_b/\ell_p=0.35$, we obtain $\ell_p=2.85 \ell_b=1.42 \sigma$.  Alternatively, we can estimate the persistence length from  $ \langle \cos \theta \rangle=0.699 $ 
  as $\ell_p=-\ell_b/\ln ( \langle \cos \theta \rangle)$ which leads to $\ell_p=2.79 \ell_b$  comparable to the estimated value from fitting the bond-bond orientational correlation functions with an exponential
  decay.  Notably,  the relation $\ell_k =2\ell_p$ valid for worm-like chains roughly holds  for semiflexible CG-PVA polymers.
  
 For larger $n$ values, similar to the MSIDs,  bond-bond correlations  deviate from exponential decay but we do not observe any sign of long-range power law decay reported for 3D melts
 of  chains with zero persistence length \cite{Wittmer2004}  and the semiflexible  polymers with a harmonic bending potential \cite{Hsu2014}. 
 A careful examination of  the bond-bond correlations of longer chains,  for which we have sufficient statistics
 (inset of  Fig. \ref{fig3}b),  shows  that $\langle  \cos \theta (n)\rangle$ becomes slightly negative for larger $n$ values with a dip around $n \approx 20$ before it decays to zero. This feature  is most likely a precursor of crystallization as
 $T=1=1.1T_c$ is slightly above the crystallization temperature. Although the negative dip in $\langle  \cos \theta (n)\rangle$ is convergent for all the three well equilibrated chain lengths,  further investigations are required to establish the existence and origin of this negative dip.  Especially, a better statistics is needed  to calculate the error bars. 
 
  In  Fig. \ref{fig4}, we have investigated the temperature dependence of $\langle \cos \theta  (n) \rangle$  for $N=200$ where the data for $T<1$ are obtained from cooling of the melt \cite{Meyer2002,SaraLMC}. As can be deduced from  Fig. \ref{fig4} at $T=0.9$ and 0.8, where the polymers are in the semicrystalline state, the negative dip is further amplified  due to local backfolding of the chains. Equilibrating the polymers  at higher temperature of $T=2$ well above $T_c$,   we confirm that their $\langle  \cos \theta (n)\rangle$ is strictly positive as expected. The  inset of  Fig. \ref{fig4} depicts a log-log plot of $\langle  \cos \theta (n)\rangle$ and shows that  beyond $n \approx 10$, bond-bond correlation exhibits a power law decay that is a signature of long-range chain self-interactions.

 
 
 %
\begin{figure*}[t]
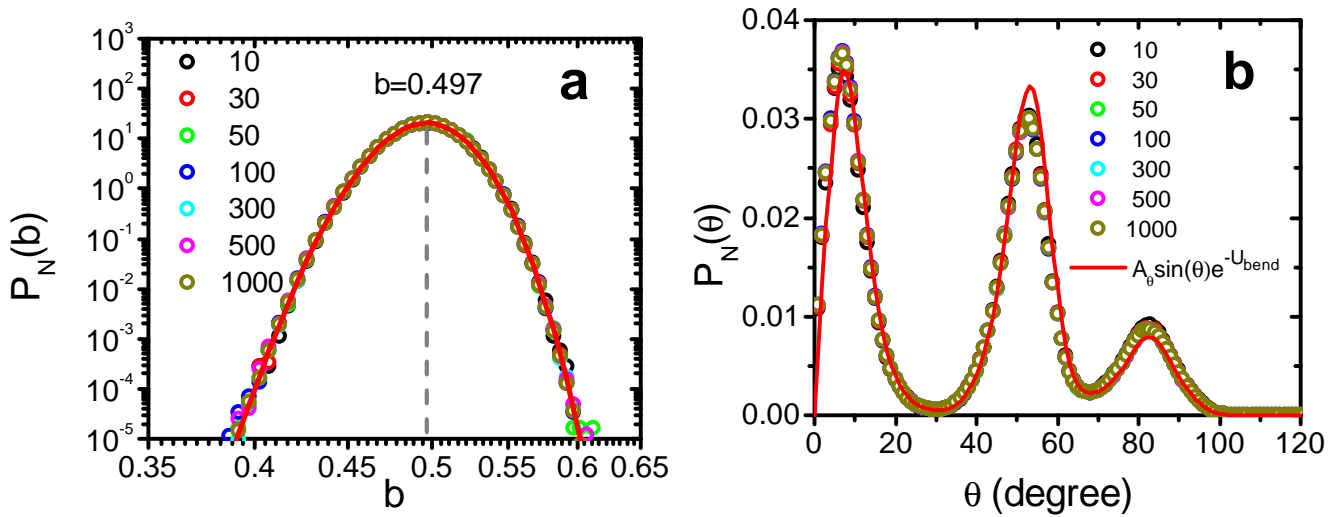

\includegraphics[width=0.5\linewidth]{fig6a.pdf}
\includegraphics[width=0.48\linewidth]{fig6b.pdf}
\caption{(a) Normalized probability distribution of the bond length $P_N (b)$ for different chain lengths as shown in the legend. The solid line depicts a fit 
$P_N (b)$ by a Gaussian distributions given in Eq. \eqref{eq:Gaussianlb} with the
 parameters $\ell_{b0}=0.49694$ and   $\sigma_{b0} = 0.01959$. (b) Normalized probability distribution of the bond angle $\theta$. The solid curve  shows the theoretical  prediction for
 the probability 
distribution of bond angles according to  $ \sin \theta \exp(-U_{bend}(\theta)/k_BT)$. }
\label{fig6} 
\end{figure*}

  To further inspect the origin of  small but systematic deviations from the FRC model, we investigate the non-Gaussian parameter of CG-PVA polymers and compare it to that of the FRC model prediction.
The non-Gaussian parameter is defined in terms of second and fourth moments of internal distances  $\alpha(n)=1/2(5-3\langle R^4(n) \rangle /  \langle R^2(n) \rangle ^2 ) $ and it  
  vanishes if the internal distances $R(n)$ are Gaussian distributed.  
    The fourth moment of internal distances $\langle  R^4 (n)\rangle$ of the FRC model depends only on  $\langle \cos \theta \rangle$ and $ \langle P_2( \cos \theta) \rangle$ 
    \cite{Honnell} where $P_2(x)=3/2 x^2-1/2$ is the second Legendre polynomial. Fig. \ref{fig5}a presents the non-Gaussian parameter of CG-PVA polymers of different chain lengths
   that is compared to that of the FRC model evaluated with $ \langle \cos \theta \rangle=0.699$ and $ \langle P_2( \cos \theta) \rangle=0.37$ extracted from simulation results. Overall, we find a good agreement between 
   the $\alpha(n)$ of CG-PVA polymers    and that of the FRC model. However, we notice small deviations from the FRC model for long chains and $n>50$.
   
   In Fig. \ref{fig5}b, we have plotted  $\langle P_2( \cos \theta(n)) \rangle$ for different chain lengths. According to the FRC theory, $\langle P_2( \cos \theta(n)) \rangle$ should also decay exponentially as 
   $ \langle P_2( \cos \theta(n)) \rangle= \langle P_2( \cos \theta) \rangle^n$ \cite{Porod1953}.
   We find that the initial decay of $ \langle P_2( \cos \theta(n)) \rangle$ up to $n=5$ is well described by the FRC model predictions. However, for larger $n$ we observe important deviations from the  FRC  model and 
   $ \langle P_2( \cos \theta(n)) \rangle$    exhibits a clear power law behavior over more than one decade of $n$.   
   The observed power law behavior is a manifestation of long-range bond-bond correlations along the chain backbone that result from incomplete screening of excluded volume interactions. 
 As noted by Meyer et. al. in the case of 2D polymer melts, it is related to  the return probability after $n$ bonds  and the local nematic ordering of nearby bonds \cite{Meyer2010}.

   To understand better the origin of this power law decay, we investigate the intrachain nematic ordering by calculating the $ \langle P_2( \cos \theta(r)) \rangle$ for all bond pairs that belong to the same chain and their midpoints
   are a distance $r$ apart.  $ \langle P_2( \cos \theta(r)) \rangle$  is almost  independent of chain length for $N>50$.  In the  inset of Fig. \ref{fig5}b, we have presented the $ \langle P_2( \cos \theta(r)) \rangle$ as a function of
   distance $r$  for $N=500$.  As can be seen, the orientational correlations oscillate and decay rapidly with $r$. This behavior shows that for a fixed curvilinear distance $n$ only  bonds which are spatially close to each other
    with separations  $r \approx 1$  contribute to $ \langle P_2( \cos \theta(n)) \rangle$. Therefore, one expects that $ \langle P_2( \cos \theta(n)) \rangle$ will be directly proportional to the return probability of monomer after  $n$ bonds
    \cite{Meyer2010}  that we denote by  $p_{\mathrm{ret}}(n)$. More precisely
        \begin{equation}
\langle P_2( \cos \theta(n)) \propto p_{\mathrm{ret}}(n) \equiv \lim_{\bR(n) \rightarrow 0} \Psi [\bR(n)] 
\label{eq:P_r(n)}
\end{equation}
   where  $\Psi [\bR(n)]$ is the probability distribution  function  of  $\bR(n)$ $\it{i. e.}$  the end-to-end vector  of all the chain segments (subchains) with  $n$ bonds.
   
   For an  ideal  self-similar chain  $\Psi [\bR(n)]$ for any subchain of size $n$ follows a Gaussian distribution of the form 
  \begin{equation}
 \Psi_{\mathrm{Gauss}} [\bR(n)]=  \left(\frac{3}{2 \pi \langle  R^2(n)\rangle} \right)^{3/2} \exp \left(-\frac{3 \bR^2(n)}{2 \langle  R^2(n) \rangle}\right)
 \label{eq:Gaussian0}
\end{equation}
where  $\langle R^2(n)\rangle \sim n^{2 \nu}$ and $\int_0^{\infty} 4 \pi R^2  \Psi (\bR)  dR =1$ \cite{Rubinstein,polymerDoi}.
Hence, the return probability scales as $p_{\mathrm{ret}}(n) \sim n^{-3 \nu} =  n^{-3/2} $. 
For semiflexible CG-PVA polymers with local correlations, we expect that $\Psi [\bR(n)]$ follows Eq. \ref{eq:Gaussian0} for $n \gg 1$.
The scaling exponent that we obtain from fitting of $\langle P_2( \cos \theta(n))$ with power law is $-1.4 \pm 0.1$,  that is not so far from the prediction for the Gaussian chains.
This small discrepancy is most likely due to insufficient statistics for large $n$ values.  To test the validity of Gaussian distribution for the internal distances we examine the behavior of  intrachain distributions in the
subsequent subsection.
 

\subsection*{Intrachain distribution functions}


 Having examined the intrachain correlations, we investigate conformational behavior of chains by extracting  the   probability distribution of the internal distances, {\it i.e.}  $P[R(n)]=4 \pi R^2(n) \Psi [\bR(n)]$ from the 
  polymer configurations.
As before, $R(n)$ denotes the distance between any pair of monomers $i$ and $j$ that are  $n \equiv |i-j|$ bonds apart. Let us first  consider the probability distribution function of bond length $P_N (b)$ corresponding to $R(n=1)$.
In Fig. \ref{fig6}a, we have shown the distribution of bond length  that is independent of the chain length $N$.    
The normalized distributions of bond length $b$ can be well described by a  Gaussian distribution of the form
\begin{equation}
P_N (b)=    \frac{  \exp[ -\frac{(b-  \ell_{b0})^2}{2  \sigma_{b0}^2} ]}{\sqrt{2\pi \sigma_{b0}^2} },
\label{eq:Gaussianlb}
\end{equation}
in which $\sigma_{b0}$ represents the  standard deviation and the peak value $\ell_{b0}=0.497$ agrees with the average bond length.

Next, we  examine the probability distributions of bond angles $P_N(\theta)$ and compare them with the form  expected from the Boltzmann distribution $P_N(\theta)= A_{\theta} \sin \theta  \exp[-\beta U_{bend} (\theta)]$ where $A_{\theta}$
is a normalization constant such that $\int_{0}^{\pi} d \theta P_N(\theta)=1$. Fig. \ref{fig6}b presents $P_N(\theta)$ obtained from accumulating the histograms of bond-angles  and the Boltzmann distribution prediction.
Overall, we find 
a good agreement between the two probability distribution functions  for all the chain lengths. 


%
\begin{figure*}[t]
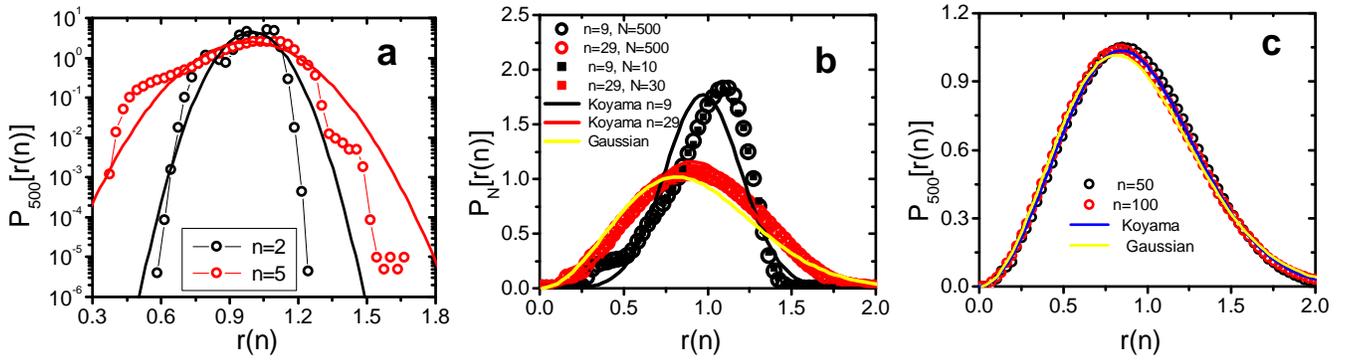

\includegraphics[width=0.335\linewidth]{fig7a.pdf}
\includegraphics[width=0.32\linewidth]{fig7b.pdf}
\includegraphics[width=0.32\linewidth]{fig7c.pdf}
\caption{Normalized probability distributions of  reduced internal  distances $r(n)=(R(n)^2/ \langle R^2(n) \rangle)^{1/2}$, $P_N[r(n)]$  
for subchains of with $n$ bonds for polymers with $N=500$ monomers. The segment sizes $n$ are given in the legends. Panel (b) also includes $P_N[r(N-1)]$
for $N=10$ and $N=30$. The solid lines represent the theoretical predictions for 
$P_N[r(n)]$  according to the Koyama distribution, Eq. \eqref{eq:Koyama}, in all the panels and the   Gaussian distribution, Eq. \eqref{eq:Gaussian}, in panels (b) and (c).  }
\label{fig7} 
\end{figure*}

Next, we focus on the normalized probability distribution of internal distances $P[R(n)]$ for   $n>1$ and compare them to the theoretical distribution functions. 
The exact segmental size  distribution functions of  semiflexible polymers for  an arbitrary $n$ are not known. However, Koyama has proposed approximate 
expressions for the probability distribution  functions of   wormlike  chain model \cite{Koyama0} that are applicable to any semiflexible polymer model  for curvilinear distances larger than the persistence segment \cite{Koyama}.
The Koyama distribution  is constructed in such a way that it reproduces the correct  second and forth moments of internal distances, {\it i.e.} $\langle R^2(n) \rangle$ and $\langle R^4(n) \rangle$ and it interpolates between 
the rigid-rod and the Gaussian coil limits \cite{Koyama}. It is found to account rather well for the site-dependence of the intrachain structure of short CG-PVA polymers \cite{Vettorel2007}.

The Koyama distribution can be expressed  in terms of scaled internal distances $r(n)=R(n)/\sqrt{\langle R^2(n) \rangle}$  as
\begin{eqnarray}
 P_{\mathrm{Koyama}}(r)=\frac{r}{\sqrt{2/3 \pi \eta (1-\eta) }}
 [ e^{ \left(-\frac{3 (r -\sqrt{\eta})^2}{2 (1-\eta)}\right)}-e^{ \left(-\frac{3 (r +\sqrt{\eta})^2}{2 (1-\eta)}\right)} ]  \nonumber \\
 \label{eq:Koyama}
 \end{eqnarray}
in which  $\eta=\sqrt{\alpha (n)}$ and $\int_{0}^{\infty} P_{\mathrm{Koyama}}(r) dr=1 $. As one would expect, at sufficiently large $n$ for which the non-Gaussian parameter $\alpha (n)$ vanishes,
the Koyama distribution becomes identical to the Gaussian distribution valid for fully flexible ideal chains. For ideal chains, the probability distribution function of the   $\bR(n)$   is given by Eq. \eqref{eq:Gaussian0}.
As a result, the  corresponding probability distribution function  for the reduced internal distances, $r(n)$,   follows from
\begin{equation}
P_{\mathrm{Gauss}}(r)=4 \pi r^2 \left(\frac{3}{2 \pi } \right)^{3/2}  \exp(-3 r^2/2 \langle r^2 \rangle)  
\label{eq:Gaussian}
\end{equation}
where $\int_{0}^{\infty} P_{\mathrm{Gauss}} (r) dr =1$.
 Particularly, one expects that the distribution functions of the end-to-end distance of sufficiently long chains should follow this distribution.

 \begin{figure*}[t]
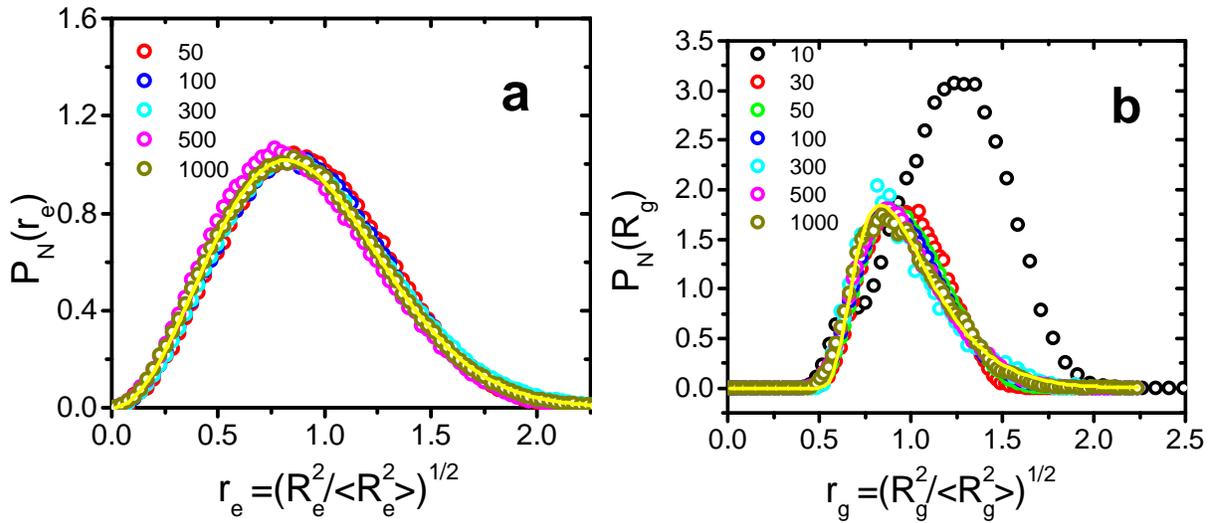

\includegraphics[width=0.44\linewidth]{fig8a.pdf}
\includegraphics[width=0.45\linewidth]{fig8b.pdf}
\caption{a) Normalized probability distribution  of  reduced end-to-end distance $r_e=(R_e^2/ \langle R_e^2 \rangle)^{1/2}$, $P_N(r_e)$  shows a very good agreement with the  Gaussian distribution  (solid line)  given in   Eq. \eqref{eq:Gaussian}. b) Normalized probability distribution of the reduced  gyration radius $r_g=(R_g^2/ \langle R_g^2 \rangle)^{1/2}$, $P_N(r_g) $. The solid line represents a fit to the theoretical prediction  according to  Eq. \eqref{eq:pdf_rg} with  parameters $\nu=1/2$, $a_1=0.125$ , $a_2=1.54$ and $A_g=7.8$. }
\label{fig8} 
\end{figure*}
Let us  first look at the distribution of internal distances for short subchains (segments). We verified the distribution of subchains does not depend on the chain length. Therefore,
we focus on  subchains of polymers with  $N=500$ monomers.
Fig. \ref{fig7}a  presents   $P[r(n)]$ for subchains  comprising of $n=2$ and 5 bonds. As can be seen, for such short segments the features of angular potential are  dominant and the Koyama distribution 
can not provide an accurate description of segmental size distribution although it agrees well with  $P[r(n)]$ in the central region of the distribution and it   captures accurately the height of the
peaks.  

Fig. \ref{fig7}b shows $P[r(n)]$ for  subchains with $n=9$ and 29 bonds that are larger than the persistence segment size. We note that the signature of bending potential is still visible for $n=9$. Thus, the Koyama distribution
does not provide a good description. For $n=29$, the Koyama distribution agrees quite well with the  $P[r(n)]$ extracted from simulations  whereas  the Gaussian distribution exhibits a poorer agreement. Additionally, we have included
the distribution of the scaled end-to-end distance of short chains with  $N=10$ and $N=30$ monomers in Fig. \ref{fig7}b. They coincide with the  distributions of subchains  with the same length. These results show that
the chain-end effects, if any, are negligible and $P[r(n)]$ only depends on $n$.
For longer subchains $n=50$ and 100, $P[r(n)]$ depicted in Fig. \ref{fig7}c  displays a perfect agreement with the Koyama distribution.  For such long subchains, the distribution functions approach to that of a Gaussian 
given by Eq. \eqref{eq:Gaussian}, as one would expect.
 
Finally, we present the normalized probability distributions of the scaled end-to-end distance $r_e$ for $N>50$ in Fig. \ref{fig8}a.
All the data   for   $P_N(r_e)$ collapse on a single master curve. We find a very good agreement between the master curves and the theoretical prediction for the $N$-independent normalized distribution function  given by Eq. \eqref{eq:Gaussian}  as demonstrated in Fig. \ref{fig8}a.

 Fig. \ref{fig8}b shows the  normalized probability distributions $P_N(r_g)$ of the scaled gyration radius $r_g= (R_g^2/ \langle R_g^2\rangle)^{1/2}$  for different chain lengths.
Similar to $P_N(r_e)$ , the $P_N(r_g)$ of chains with $N \ge 50$   collapse on a single master curve.
The exact expression for the probability distribution of the gyration radius  is more complicated and does not have a compact form \cite{pdfRg}. However,  the  formula suggested by Lhuillier \cite{Lhuillier} for polymer chains under good solvent conditions   is found to  provide  a good approximation for ideal chains   \cite{pdfRg1,Hsu2014,Kremer2016} too. The Lhuillier formula for the scaled 
gyration radius $r_g$  in $d$-dimensions reduces to
\begin{equation}
P_N(r_g)= A_g \exp \left( - a_1 r_g^{-\alpha d} -a_2 r_g^{\delta} \right)
\label{eq:pdf_rg}
\end{equation}
in which  the exponents $\alpha$ and $\delta$ are related to the space dimension $d$ and the Flory exponent $\nu$ by $\alpha=(\nu d-1)^{-1}$ and $\delta=(1-\nu)^{-1}$.  
$a_1$ and $a_2$ are system-dependent non-universal constants and $A_g$ is a normalization constant such that  $\int_{0}^{\infty} P_N(r_g) dr_g=1$.
We find that  the data of $P_N(r_g)$ of CG-PVA polymers can be well fitted by the $N$-independent normalized distribution function  given by  Eq. \eqref{eq:pdf_rg}  as
plotted  in Fig. \ref{fig8}b.   Having investigated the conformational properties of CG-PVA polymers,  we focus on their structural properties
in the Fourier space in the next subsection \cite{Vettorel2007,Meyer2010}.

\subsection*{ Form factor and  structure  factor}
\begin{figure*}[t]
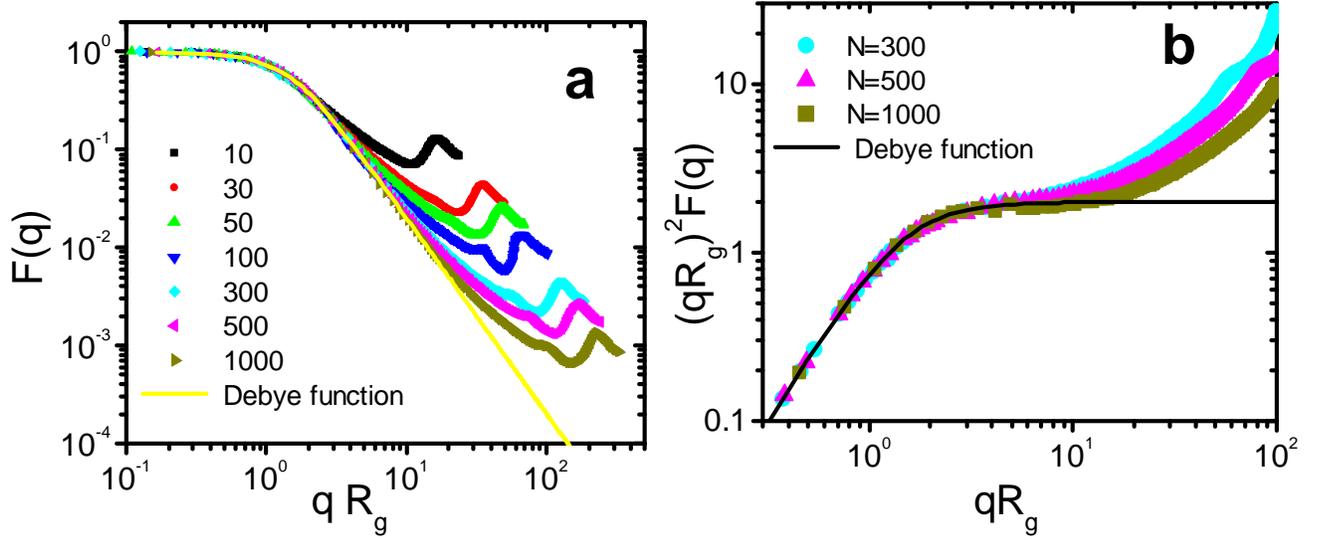

\includegraphics[width=0.48\linewidth]{fig9a.pdf}
\includegraphics[width=0.48\linewidth]{fig9b.pdf}
\caption{
(a) Form factors $F(q)$ plotted against $qR_g(N)$ for different chain lengths. (b) The form factor of the longer chains $N=300,500$ and 1000  in a Kratky plot  {\it i.e.} $(qR_g)^2 F(q)$ 
versus $qR_g$. The $R_g$ values for different chain lengths are given in Table I.
   }
\label{fig9} 
\end{figure*}

A  common way to characterize the structural  properties of polymer melts is to explore their  structure factor that can be measured directly in the scattering experiments.
The structure factor encompasses the information about spatial correlations between the monomers via Fourier transform of  density-density correlation functions.    For spatially homogeneous and isotropic 
systems such as polymer melts at equilibrium, the static structure factor only depends on the modulus $q$ of the wave vector.
The static structure factor $S(q)$  measured in scattering experiments of amorphous melts  is  often spherically averaged over all the wave vectors $\mathbf{q}$ with the  same modulus $q$.  This quantity can be
computed as 
\begin{equation}
S(q)=\frac{1}{N n_c} \sum_{i,j=1}^{n_c} \sum_{n,m=1}^{N} \langle \exp \left[-i \mathbf{q} \cdot (\mathbf{r}_{i,n} -\mathbf{r}_{j,m}) \right] \rangle
\label{eq:SQ0}
 \end{equation}
where the angular brackets represent  averaging over all the wave vectors with the same modulus and all the melt configurations.

$S(q)$ given in Eq. \eqref{eq:SQ0} encompasses scattering from all the  monomer pairs. It can be split into intrachain and interchain  contributions
\begin{equation}
S(q)= S_c(q)+ \rho_m h(q)
\label{eq:SQ}
 \end{equation}
where $\rho_m=Nn_c/V$ (V volume of the simulation box) is the monomer density and 
\begin{equation}
S_c(q)= \frac{1}{N n_c} \sum_{i=1}^{n_c} \sum_{n,m=1}^{N} \langle \exp \left[-i \mathbf{q} \cdot (\mathbf{r}_{i,n} -\mathbf{r}_{i,m}) \right] \rangle\label{eq:SQ1}
 \end{equation}
includes the contributions from intrachain pair correlations and it is called  intrachain or single chain structure factor. Equivalently,  $F(q)=S_c(q)/N$ known as the form factor \cite{Rubinstein} is used to quantity
the intrachain correlations in the Fourier space. The interchain contribution is given by $h(q)$ that is defined as the Fourier transform of intermolecular pair correlation function \cite{Hansen} as
\begin{equation}
h(q)= \frac{V}{(N n_c)^2} \sum_{i\neq j}^{n_c} \sum_{n,m=1}^{N} \langle \exp \left[-i \mathbf{q} \cdot (\mathbf{r}_{i,n} -\mathbf{r}_{j,m}) \right] \rangle\label{eq:SQ2}.
 \end{equation}
\begin{figure}[t]
 \includegraphics[width=0.98\linewidth]{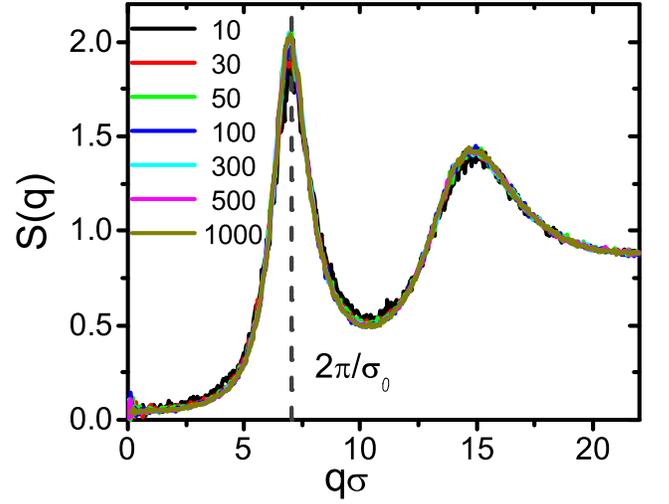}
\caption{The total structure factor versus $q$ for different chain lengths of CG-PVA polymers in the melt state.}
\label{fig10} 
\end{figure}
\begin{figure*}[t]
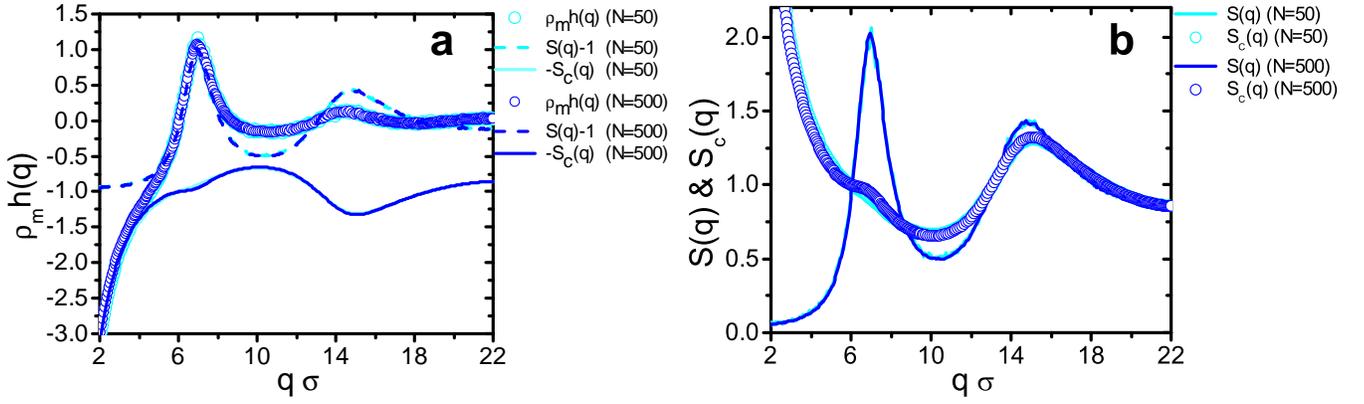

\includegraphics[width=0.50\linewidth]{fig11a.pdf}
\includegraphics[width=0.49\linewidth]{fig11b.pdf}
\caption{Comparison of (a) the interchain structure factor $\rho_m h(q)$ and $S(q)-1$  (b) the intrachin structure factor  $S_c(q)$ and total structure factor  of CG-PVA polymer melts with $N=50$ and 500.  }
\label{fig11} 
\end{figure*}
  We present the  behavior of the form and  structure factors for different chain lengths. We first focus on the form factor as depicted in   Fig. \ref{fig9}. 
   The form factor of Gaussian chains,  known as  the  Debye function, is described by \cite{Rubinstein}
  \begin{equation}
F_{Debye}(q)= \frac{2}{Q^2}  \left[ \exp(-Q)+Q-1   \right]   \quad \text{with} \quad  Q=q^2 \langle R_g^2\rangle
\label{eq:SQDebye}
 \end{equation}
 In order to compare the behavior of CG-PVA  polymers in the melt state with that of ideal Gaussian chains, in  Fig. \ref{fig9}a, we have plotted the form factor of CG-PVA chains 
 and the Debye function versus $qR_g$. 
 For all the chain lengths, we  observe deviations from  the ideal polymer behavior at high $q$ values. The onset of deviations  shifts progressively
 to larger wave vectors for longer chains. For the longest chains  $N \ge 300$, we  have also  presented the form factors $F(q)$  in a Kratky-plot in Fig. \ref{fig9}b.
 This plot confirms the existence of a  \emph{Kratky plateau} in the scale free regime that extends up to  $qR_g \approx20$ for  $N=1000$.
  The deviations at larger $q$ values reflect the 
 underlying form of the  bond angle potential  $U_{bend} (\theta)$ that dominates the behavior of the form factor at length scales smaller or  comparable to the Kuhn length.

  Next, we present the structure factor of different chain lengths in Fig. \ref{fig10}. As we notice, the $S(q)$ of all the polymer melts displays the characteristic features  of the liquid-state. 
  We find a very weak dependence on the chain
  length; for $N> 50$, the $S(q)$ of various chain lengths  are identical. We notice several important features in the structure factors. 
    First, the structure factor at low $q$ is very small. By virtue of compressibility equation that relates the isothermal compressibility $\kappa_T$ to the structure of the liquid,
  {\it i.e.} $\lim_{q\rightarrow 0} S(q)=\rho_m k_b T \kappa_T $, we conclude that  the polymer melts are almost incompressible. 
 Second, the first peak of $S(q)$ at $q^*$ characterizes the packing of monomers in the first nearest neighbor shell. The value of $q^*$ nearly agrees  with $2 \pi/ \sigma_0$ reflecting that the first peak of
 $S(q)$ is dominated by interchain contributions.

To gain more insight into the interchain correlations of CG-PVA polymers, we compare $\rho_m h(q)$ with that of simple liquids with no internal structure. For such a simple liquid, we have $S_c(q)=1$, hence
$\rho_m h(q)=S(q)-1$ \cite{Hansen}. Fig. \ref{fig11}a shows both $\rho_m h(q)$ and $S(q)-1$ for two chain lengths $N=50$ and $N=500$. We find that in the region near the first peak $\rho_m h(q)$ and $S(q)-1$ coincide confirming that the
peak at $q=q^*$ is totally determined by the interchain correlations.  In Fig. \ref{fig11}a, we have also included $-S_c(q)$. We note that for very low wave vectors beyond the peak region, $\rho_m h(q)$ closely agrees with $-S_c(q)$.
This behavior shows that the correlation between monomers of different chains decreases with increasing distance.  This decrease is concomitant by the increase of $S_c(q)$ at low $q$ values such that the sum of both intrachain and interchain
contributions yields a small finite value for $S(q)$ as $q \rightarrow 0$. 

In the other extreme of $q \gg q^*$,    $\rho_m h(q)$  deviates from  $S(q)-1$   as the large $q$ behavior of structure factor is fully determined by
the intrachain correlations due to the  correlation hole effect  \cite{polymerDeGennes,Vettorel2007}. The correlation hole effect leads  to a decreased probability
of finding a monomer of another chain in the pervaded volume of a particular chain.
To illustrate this point, in  Fig. \ref{fig11}b, we have shown  $S_c(q)$ and $S(q)$ for  $N=50$ and $N=500$  in the same plot. We see that the large-$q$ behavior is entirely dominated by intrachain
contributions.  These observations are in agreement  with the prior investigations for short chain lengths
$ 10\le N \le 100$ \cite{Vettorel2007}. 

 \begin{figure}[h]
\includegraphics[width=0.98\linewidth]{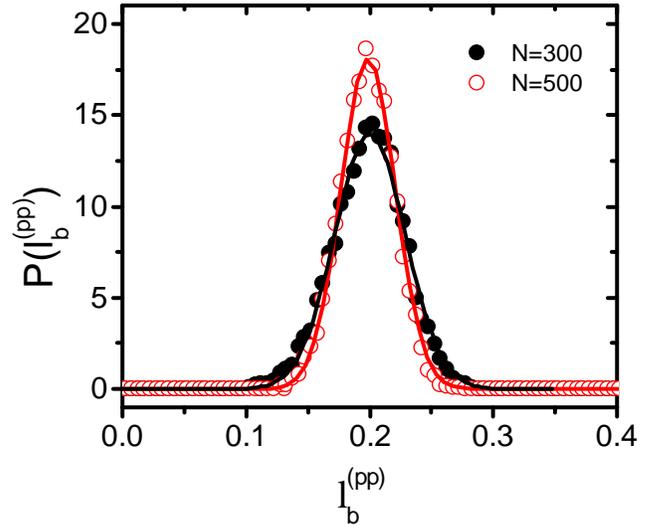}
\caption{Normalized probability distributions of  bond length of the primitive paths $P_N(\ell_b^{(pp)})$ for $N=300$ and 500. The solid lines show  fits with 
the Gaussian distribution given in Eq. \eqref{eq:Gaussian1} with $\langle \ell_b^{(pp)}\rangle=0.20 \pm 0.002$, $\sigma_{(N=300)}=0.0287 \pm 0.001$ and $\sigma_{(N=500)}=0.022 \pm 0.001$. }
\label{fig12} 
\end{figure}
\begin{figure*}[t]
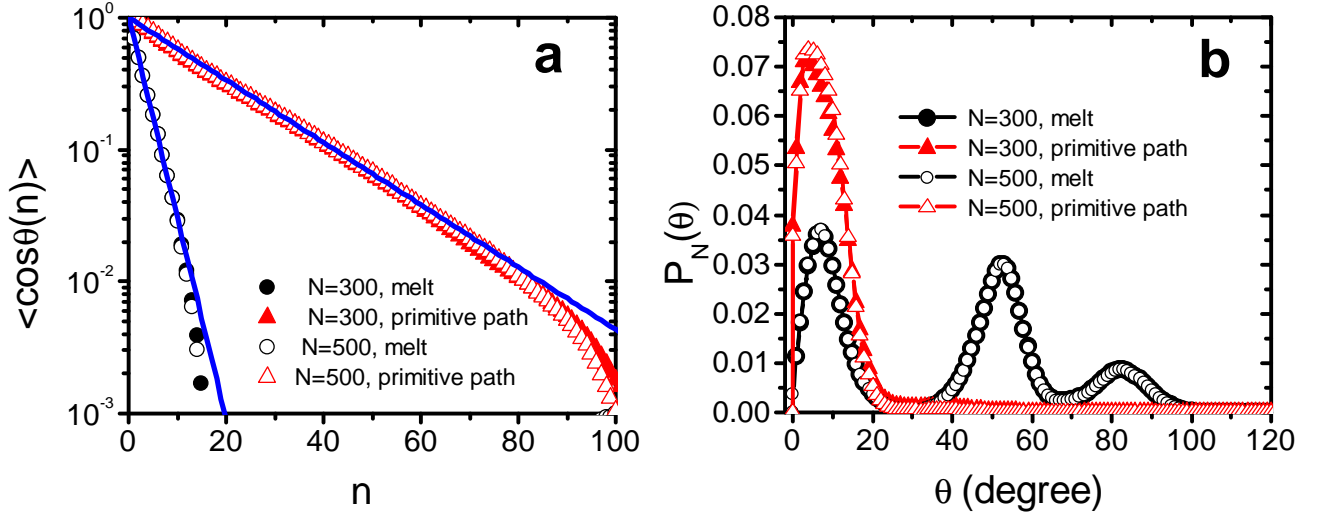

\includegraphics[width=0.48\linewidth]{fig13a.pdf}
\includegraphics[width=0.48\linewidth]{fig13b.pdf}
\caption{(a) The bond-bond orientational correlation function $ \langle \cos \theta  (n) \rangle$ versus $n$.  The straight  lines show the fits  with an exponential  functions with 
the form $\exp(-\ell_b/\ell_p n$) with $\ell_b/\ell_p=0.35 \pm 0.01$ and $\exp(-\langle \ell_b^{(pp)}\rangle/\ell_p^{(pp)} n)$ with $\langle \ell_b^{(pp)}\rangle/\ell_p^{(pp)}=0.0544 \pm 0.005$ .  (b) Normalized probability distribution of the angle $\theta$ between any two consecutive bonds for  primitive paths and original conformations of  CG-PVA polymer mets with $N=300$ and 500.  }
\label{fig13} 
\end{figure*}

\section*{Primitive path analysis and entanglement statistics}
%

%
Having investigated conformational and structural features of CG-PVA polymer melts, we  focus on their topological characteristics, {\it i.e.} interchain entanglements. 
Entanglements stem from topological constraints due to the chain  connectivity and uncrossability that restrict the  movements of chains at the intermediate time and length scales.  As first noted by Edwards 
\cite{Edwardstube}, the presence of neighboring strands in a dense polymer melt effectively confines  a single polymer strand to a tube-like region.  The  centerline  of such a tube is known as the primitive path (PP).   
 A practical and powerful method for characterizing the entanglements is primitive path analysis (PPA). Such an analysis provides us with an operational definition of primitive path and allows   to investigate  statistics of chain  entanglements.

There exists  a couple of variants of  PPA  in the literature \cite{Kroger,Hoy2009,PPAEveraers} that   are all similar in spirit. Here, we implement the  PPA method proposed by Everaers et al. \cite{PPAEveraers}
that  identifies the primitive path of each polymer chain in
a melt based on the concept of Edwards tube model \cite{Edwardstube}.
The primitive path is defined as the shortest path between the chains ends 
that can be  reached from the initial conformations of polymers without crossing other chains.  In this analysis, topologies of chains are conserved, and chains are assumed to follow
random walks along their primitive paths. Therefore, the primitive path is a random walk with the same mean square end-to-end distance $ \langle R_e^2 \rangle= \langle R_e^2 \rangle^{(pp)}$  but  shorter 
bond  length  $\ell_b^{(pp)}$ and contour length $L^{(pp)}=(N-1)\ell_b^{(pp)}$.

In practice,  by extracting the average bond length of the primitive paths $\langle \ell_b^{(pp)} \rangle=1/ (N-1) \langle \sum_{i=1}^{N-1}|\mathbf{r}_{i+1}-\mathbf{r}_i|\rangle $, we can determine
all the other desired quantities.  In particular, the Kuhn 
length of primitive path $\ell_K^{(pp)}$ is obtained as
\begin{equation}
 \ell_K^{(pp)}=\frac{\langle R_e^2 \rangle}{\langle L^{(pp)} \rangle}=\frac{ \langle R_e^2 \rangle}{ (N-1) \langle \ell_b^{(pp)} \rangle}.
\label{eq:KuhnPP}
  \end{equation}
The so-called entanglement length $N_e$, defined as the average number of monomers in the Kuhn segment of the 
primitive path,  follows from 
\begin{equation}
N_{e}=  \frac{\ell_K^{(pp)}}{\langle \ell_b^{(pp)}\rangle}= \frac{\langle R_e^2 \rangle}{(N-1) \langle \ell_b^{(pp)}  \rangle^2 }.
 \end{equation}

Operationally,  we  obtain the primitive paths of polymers in a melt  by slowly cooling the system toward $T = 0$ while  the two chain ends are kept fixed. During this procedure, the intrachain excluded volume interactions and  bond angle 
potential are switched off.  The system is then equilibrated using a conjugate gradient algorithm in order to minimize its potential energy and reach a local minimum. We perform primitive path analysis for
the two longest chain lengths that are fully equilibrated, {\it i.e.} $N=300$ and 500 as it is known that poor equilibration affects the entanglement length \cite{Hoy2009}.

We first examine the probability distributions of  the bond lengths of  the primitive paths, {\it i.e.} $P_N(\ell_b^{(pp)})$ as presented in Fig. \ref{fig12}. The distributions of bond lengths of the primitive paths  are chain length dependent 
but both are centered at  $\ell_b^{(pp)}=0.20$. Furthermore, the primitive path bond length 
 fluctuations  are considerably larger than those of their original paths. The normalized distributions of primitive path bond length $\ell_b^{(pp)}$ can be well described by a Gaussian distribution  of the form 
\begin{equation}
P_N(\ell_b^{(pp)})= \frac{1}{\sqrt{2\pi \sigma_{N}^2} }\exp \left(\frac{(\ell_b^{(pp)}-\langle \ell_b^{(pp)}\rangle)^2}{2  \sigma_N^2} \right).
\label{eq:Gaussian1}
\end{equation}
where $\sigma_N=\langle {\ell_b^{(pp)}}^2\rangle-\langle \ell_b^{(pp)}\rangle^2$ presents the  $N$-dependent standard deviation of $\ell_b^{(pp)}$.
\begin{figure}[t]
\includegraphics[width=0.98\linewidth]{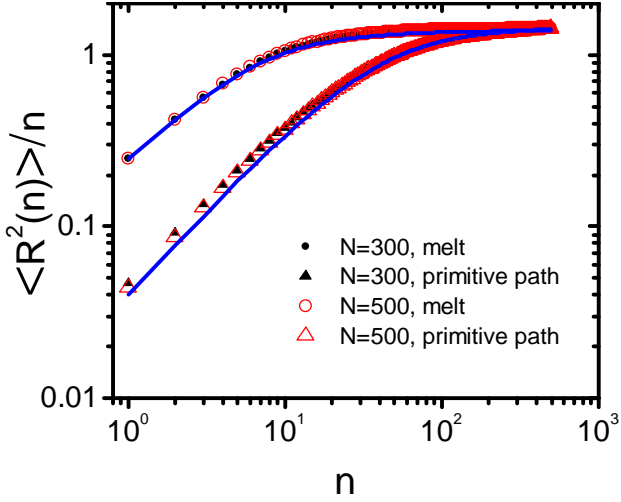}
\caption{Rescaled mean square internal distance, $\langle R^2(n) \rangle/ n  \langle b^2 \rangle $  of the original and  primitive paths of  polymer melts with chain lengths $N=300$ and 500. The solid lines show the theoretical prediction of  the equivalent freely rotating
chain (FRC)  with $ \langle \cos \theta \rangle=0.6985 $ and   $ \langle \cos \theta \rangle^{(pp)}=0.947$.  }
\label{fig14} 
\end{figure}

Next, we  investigate the statistical features of  bond angles of the primitive paths. Fig. \ref{fig13}a presents the  bond-bond orientational correlation function $ \langle \cos \theta  (n) \rangle$ as a function of
internal distance  $n$ for the primitive paths of chain lengths $N=300$ and 500.  For comparison, we have also shown the $ \langle \cos \theta  (n) \rangle$ of the original polymer conformations.
Similar to the original polymer conformations, the initial decay of  $ \langle \cos \theta  (n) \rangle$  for $10< n < 80$ can be well described by an exponential decay. However, at
short scales $n<10$, bonds are slightly stretched out because of the constraints of fixed chain ends during minimization of primitive path length.
Assuming an exponential function of the form $\exp(-n \langle \ell_b^{(pp)}\rangle/\ell_p^{(pp)} )$, we can extract the persistence length of the primitive path $\ell_p^{(pp)}$.
 From the fit values, we find  $\ell_p^{(pp)}=19 \langle \ell_b^{(pp)}\rangle=3.80 \sigma$ that is considerably larger than the persistence length of the original conformations $\ell_p=1.42 \sigma$.
 
 We also examine the normalized probability distributions of bond angles $\theta$ of the primitive paths as displayed in Fig. \ref{fig13}b.
Unlike the bond angle distributions of the original chain conformations, the bond angles of the primitive paths is unimodal with its peak centered around $\theta=4.5^{\circ}$. Furthermore, the range of angles
shrinks from $[0^{\circ},100^{\circ}]$ for the original paths to $[0^{\circ},20^{\circ}]$ for the primitive paths reflecting that the primitive paths are mainly in stretched conformations.
 \begin{figure}[t]
\includegraphics[width=0.98\linewidth]{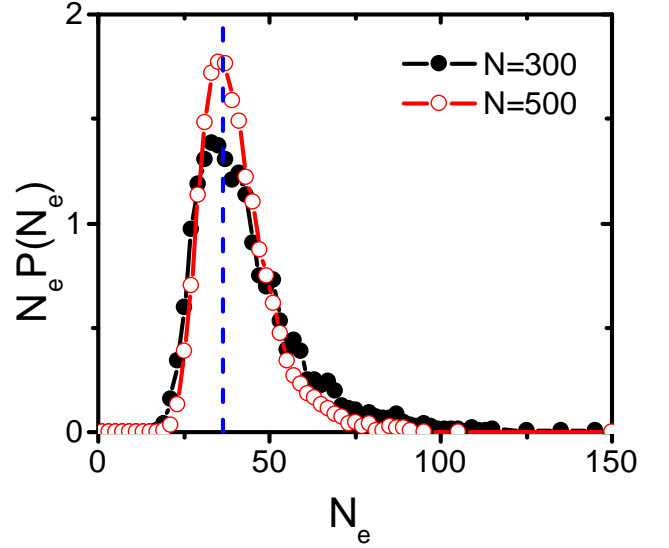}
\caption{ Normalized probability distributions of the entanglement lengths $ P(N_e)$ times the entanglement length $N_e$.  
 The dashed line shows the position of average entanglement length $\langle N_e \rangle = 36.5 $ that coincides roughly with the peak position of $N_e P(N_e)$.
  }
\label{fig15} 
\end{figure}

To explore the intrachain correlations of the primitive paths, we have plotted  the mean square internal distances $\langle R^2(n)\rangle/n$ of
the original and  primitive paths in Fig. \ref{fig14}. As expected the values  of $\langle R^2(n)\rangle/n$ for
both paths approach the same value with increasing $n$, since the  chains endpoints  during the primitive path analysis are held fixed. We find that results of $\langle R^2(n) \rangle$
for the primitive path can still be relatively well described by the generalized FRC model provided that we use  $ \langle \cos \theta \rangle^{(pp)}= \exp(-\langle \ell_b^{(pp)}\rangle/\ell_p^{(pp)})=0.947 $ extracted
from bond-bond orientational correlations. Having confirmed that the mean end-to-end distance of the primitive paths remain  identical to those of the original chains, we obtain $\ell_K^{(pp)}=7.1 \pm 0.05 \sigma$   for  the Kuhn length of 
the primitive path. We note that the $\ell_K^{(pp)}$ is  larger than the Kuhn length of polymers $\ell_K=2.93$ $\sigma$.

Subsequently, we  acquire the distribution of entanglement length $P (N_e)$ as presented in Fig. \ref{fig15}a. We notice that
$P (N_e)$  has a narrow distribution and  presents a weak dependence on $N$ possibly resulting from the finite size of the chains. 
Our estimated value of the average entanglement length  is $\langle N_e \rangle = 36.2 $ for $N=300$  and $\langle N_e \rangle = 36.6 $ for $N=500$. These results suggest that
we are rather close to the asymptotic value of entanglement length $N_e^{\infty}$.
 We have also plotted $N_e P (N_e)$ in Fig. \ref{fig15}b and we find that the position of the peak of $N_e P (N_e)$  coincides with our estimated value of  $\langle N_e \rangle \approx 36.5$.
This observation is in agreement with the PPA analysis results for the Kremer-Grest (FENE) model \cite{Kremer2016}.

 \section*{Dynamic scaling of monomer motion}

To compare the dynamic behavior of  polymer melts with the  predictions of  the Rouse and reptation theories \cite{polymerDoi}, we measure
  the mean square displacement (MSD) of monomers for  short ($N<N_e$) and  long chains ($N>N_e$). The quantities  often used to characterize the segmental motion of
polymer chains in a melt are listed below. \\
i) the mean square displacement of inner monomers  
\begin{equation}
 g_1(t)=\frac{1}{n_c (N/2+1)}  \sum_{i=1}^{n_c} \sum_{j=N/4}^{3N/4} \langle [\mathbf{r}_{i,j}(t)-\mathbf{r}_{i,j}(0)]^2 \rangle 
\end{equation}
where only the monomers in the central region of a chain are considered to suppress the fluctuations caused by the chain ends.  \\
ii) the mean square displacement of inner monomers with respect to
the corresponding center of mass (c.m.)  obtained as:
\begin{equation}
 g_2(t)=\frac{1}{n_c N}  \sum_{i=1}^{n_c} \sum_{j=1}^{N} \langle [(\mathbf{r}_{i,j}(t)-\mathbf{r}_{i,cm}(t))-(\mathbf{r}_{i,j}(0)-\mathbf{r}_{i,cm}(0))]^2 \rangle 
\end{equation}
  iii) the mean square displacement of the center of mass of chains defined as:
\begin{equation}
 g_3(t)=\frac{1}{n_c}  \sum_{i=1}^{n_c}  \langle [\mathbf{r}_{i,cm}(t))-\mathbf{r}_{i,cm}(0))]^2 \rangle 
\end{equation}

 For short chains the topological constraints do not play a dominant
role. Hence,  all the interchain interactions can be absorbed into a monomeric friction and a coupling to a heat
bath. The dynamics of the chain can then be described by a Langevin
equation with noise and the constraint that the monomers
are connected to form a chain. This description is known as the Rouse random walk model \cite{polymerDoi,Rubinstein}.
Beyond the microscopic timescale $\tau_0$, the Rouse model predicts that both  $ g_1(t)$ and $g_2(t)$  scale as $\sim t^{1/2}$ for timescales $ \tau_0< t < \tau_R $
in which  $\tau_R = \tau_0 N^2$ is known as  the Rouse time. It corresponds to the longest relaxation time of  polymers.  At longer times corresponding to
$t >\tau_R $, $ g_1(t)$ is expected to follow the Fickian diffusion and scale as $t$ whereas $g_2(t)$ is predicted to exhibit a plateau.
  The mean-square displacement of chain's center of mass is predicted to be diffusive $ g_3(t)=6 D_{\text{cm}} t$ at all  the times $t > \tau_0 $ 
  with the diffusion coefficient scaling as $D_{\text{cm}} \sim  1/ N$.

\begin{figure}[t]
\includegraphics[width=0.99\linewidth]{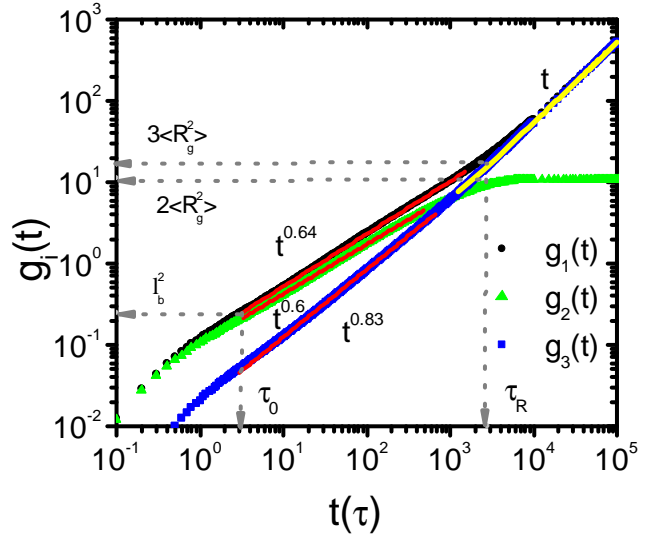}
\caption{Mean square displacement (MSD) of inner monomers $g_1(t)$, MSD relative to the center of mass of
  each chain  $g_2(t)$ and the MSD of center of  mass $g_3(t)$ of polymers of length  $N = 30$ as a function of time. The solid lines show the best fits with power law for timescales smaller and larger than the Rouse time $\tau_R$.  
shown by solid lines for comparison. }
\label{fig16} 
\end{figure}

\begin{figure*}[t]
\includegraphics[width=0.33\linewidth]{fig17a.pdf}
\includegraphics[width=0.33\linewidth]{fig17b.pdf}
\includegraphics[width=0.325\linewidth]{fig17c.pdf}
\caption{ a) Mean square displacement of inner monomers $g_1(t)$ b)mean square displacement of  monomers with respect to the center of mass of
the each chain  $g_2(t)$ and c) mean square displacement of center of mass $g_3(t)$ (green) and $g_1(t)$ (black) plotted versus time  for $N = 300$. }
\label{fig17} 
\end{figure*}
\begin{figure*}[t]
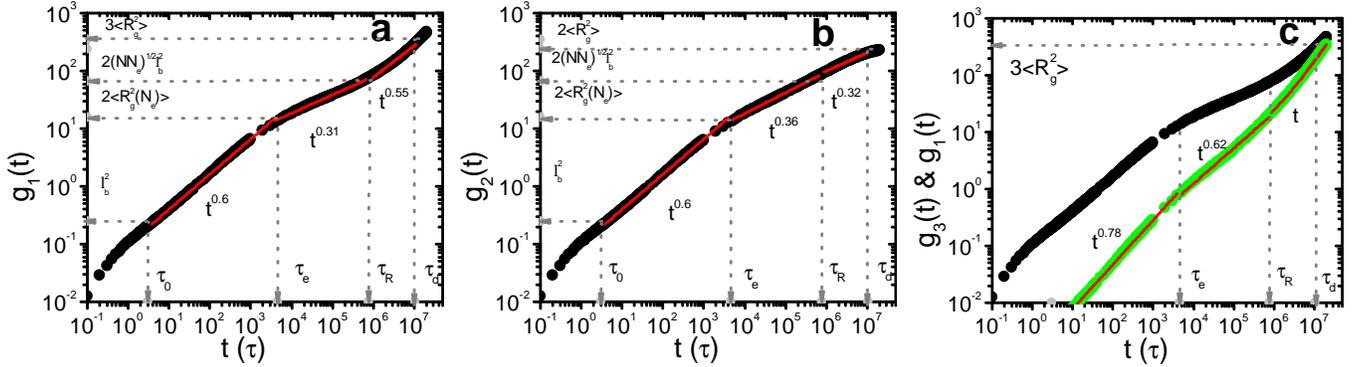

\includegraphics[width=0.33\linewidth]{fig18a.pdf}
\includegraphics[width=0.33\linewidth]{fig18b.pdf}
\includegraphics[width=0.325\linewidth]{fig18c.pdf}
\caption{ a) Mean square displacement of inner monomers $g_1(t)$ b)mean square displacement of  monomers with respect to the center of mass of
the each chain  $g_2(t)$ and c) mean square displacement of center of mass $g_3(t)$ (green) and $g_1(t)$ (black) plotted versus time  for $N = 500$. }
\label{fig18} 
\end{figure*}

Fig. \ref{fig16} presents the results  of computed $g_1(t)$, $g_2(t)$ and $g_3(t)$  for  short non-entangled polymers of chain length $N=30$.
   The microscopic timescale  $\tau_0 \approx 3.1 \tau $   is obtained as the time for which $g_1(t)= \ell_b^2$ where $\tau$ is the Lennard-Jones (LJ) time unit. The Rouse time is estimated as $\tau_R = \tau_0 N ^2 \approx 2.79 \times 10^3 \tau$.  The dotted vertical lines mark the microscopic and Rouse timescales.  As can be seen the Rouse timescale  agrees well with the crossover points between two different scaling regimes.
  The horizontal dotted lines in Fig.~\ref{fig16} depict the  values of  $g_1(t)$ and $g_2(t)$ at $t=\tau_R$. We find that $g_1(\tau_R)\approx  \langle 3⟨R^2_g(N)\rangle $ and $g_2(\tau_R)\approx  \langle 2⟨R^2_g(N)\rangle$. The  solid lines show the best fits   with  power law. For timescales smaller than a Rouse time,  $g_1(t)$ and $g_2(t)$  exhibit a subdiffusive behavior with exponents $0.64$ and  $0.6$  which are slightly higher than 
  the Rouse random walk model scaling prediction of $t^{1/2}$.   The observed disagreement most likely originates from the finite persistence length of polymers. Additionally, at this regime $g_2(t)$  is dominated by the crossover to the plateau. For chains with a persistence length smaller than their contour length, one expects that the monomer mean-square displacement scaling exponents to interpolate between those of the Rouse models for  polymers with zero persistence length, {\it i.e.} $t^{1/2}$  \cite{polymerDoi,Rubinstein}  and semiflexible worm-like polymers $t^{3/4}$ \cite{Barkema2014}.

 The mean square displacement of center of mass of polymers also deviates from the Rouse theory. At intermediate times  $  \tau_0 < t < 200$ $\tau$, it  displays  an apparent power law of the form $g_3(t) \sim t^{0.83}$ before a crossover to the expected diffusive regime. The observed subdiffusive behavior is predicted by the recent theories ~\cite{Farago1,Farago2} that account for the viscoelastic hydrodynamic interactions.  These theories attribute the anomalous behavior of $g_3(t)$ in the transient regime to incomplete screening of hydrodynamic interactions beyond the monomer length combined with the time-dependent viscoelastic relaxation of the melt. This leads to a subdiffusive motion which is not described by a pure power law; an effective exponent should decrease with chain length, but asymptotic behavior is reached extremely slowly \cite{Farago1,Farago2}. Inspection of Fig. \ref{fig16} shows that the data of $g_3$ indeed exhibits a curvature with respect to the fitted power law and the effecctive exponent of 0.83 is  quite close to 1.  The  asymptotic dynamic scaling exponent for  fully flexible chains with the Langevin dynamics is predicted to be $3/4 $. Considering the finite persistence length  and chain size,  this value is not surprising.

  In long entangled polymer melts, each chain is expected to move
back and forth (reptation) inside an imaginary tube with  a diameter $d_T  \approx N_e^{1/2}  \ell_b  \sim   R_e (N_e)$  and a contour  
length $L^{(pp)}$ around the so-called primitive path. Such a dynamical tube-like confinement affects the segmental motion of entangled polymers.  According to the reptation theory,  the dynamic scaling behavior of MSD  should exhibit a crossover behavior at several
timescales, the microscopic timescale $\tau_0$, the entanglement time $\tau_e \sim \tau_0 N_e^2$, the Rouse time  $\tau_R \sim \tau_0 N^2$, and the disentanglement
time   $\tau_d \sim \tau_0 N^3/N_e$ for long chains with $N \gg N_e$. 
The  scaling predictions of reptation theory  for  various time regimes are given by \cite{polymerDoi,KGmodel}:
 \begin{equation}
    g_1(t) \sim
    \begin{cases}
      t^{1/2}, &  \tau_0< t < \tau_e \\
     t^{1/4}, &  \tau_e< t < \tau_R\\
     t^{1/2}, &  \tau_R< t < \tau_d \\
      t , &  \tau_d< t ,
       \end{cases}
       \label{eq:g1}
  \end{equation}
  Likewise, $ g_2(t)$  is predicted to show the same regimes for $t<\tau_d$ but to go to a plateau with a value of $2 \langle R_g^2 \rangle$ for $t>  \tau_d$.
  The reptation theory predicts the following scaling behavior for the polymer  center of mass  mean square displacement  
  \begin{equation}
    g_3(t) \sim
    \begin{cases}
      t, & t < \tau_e \\
      t^{1/2}, &  \tau_e< t < \tau_R\\
      t , &  \tau_R< t   \\
    \end{cases}.
     \label{eq:g3}
  \end{equation}

  To examine these scaling predictions, we compute $g_1(t)$, $g_2(t)$ and $g_3(t)$ for   chain lengths $N=300$ and $N=500$   up  to $t \approx 2 \times 10^7 \tau$. These chain lengths include   on the average 8  and 14 entanglement lengths.
 The results of computed $g_1(t)$, $g_2(t)$ and $g_3(t)$  are presented in  Figs. \ref{fig17} and  \ref{fig18}. 
 For both chain lengths, similar to the predictions of reptation theory, we observe crossovers between several scaling regimes. 
 We first estimate the predicted crossover timescales.
The microscopic timescale  $\tau_0 \approx 3.1 \tau $ is identical to that for short chains.
 The entanglement time is estimated as the Rouse time of the chain segment between two consecutive entanglements;  $\tau_e = \tau_0 N_e^2 \approx 4.13 \times 10^3 \tau$. Likewise the Rouse time is obtained as $\tau_R(N) = \tau_0 N ^2$ leading to $\tau_R(300)\approx 2.79 \times 10^5 \tau$ and $\tau_R(500) \approx 3.28 \times 10^6 \tau$. The longest relaxation time   determined as $\tau_d(N) = \tau_0 N^3/N_e$ corresponds to $\tau_d(300)\approx 2.29 \times 10^6 \tau$ and
 $\tau_d(500)\approx 1.08 \times 10^7 \tau$. The vertical lines in Figs. \ref{fig17} and  \ref{fig18} mark these  timescales. 
 As can be noticed these timescales agree well with the crossover points between  different scaling regimes  for both chain lengths. 
 
  The horizontal dotted lines in  panels a and b   of Figs. \ref{fig17} and  \ref{fig18} present the  values of  $g_1(t)$ and $g_2(t)$ at the corresponding
timescales according to the reptation theory predictions. At $t=\tau_e$ where a Rouse chain of $N_e$
monomers is relaxed, the monomers displacement should be proportional to the mean squared diameter of the confining tube diameter $d_T$. Assuming that $d_T^2 \approx R_e^2(N_e)$ and  the Gaussian picture for the tube, we expect that 
$g_1(\tau_e)\sim d_T^2/3 \approx 2R_g^2(N_e) $.  For $t>\tau_e$ where the entanglement effects set in,  monomer  motions are restricted to movements along the contour of  the confining tube with a contour length $L_T=d_T N/N_e$ until reaching the Rouse timescale $\tau_R (N)$. Since the tube itself is a random walk with a step length $d_T$, the displacement of a monomer at $t=\tau_R$ is  $g_1(\tau_R)\sim d_T^2 (N/N_e)^{1/2} \sim   (N N_e)^{1/2} \ell_b^2$. For $ \tau_R < t < \tau_d$,  the dynamics of $g_1(t)$ is expected to crossover to a second $t^{1/2}$ regime which corresponds to the diffusion of  the whole chain  inside  the tube-like region.   After reaching the disentanglement time (reptation
time) a chain has moved a distance comparable to its own size $g_1(\tau_d)\approx  g_3(\tau_d) \approx \langle R^2_e(N)\rangle/2= 3\langle R^2_g(N)\rangle$. The initial tube is completely destroyed and a new tube-like regime reappears. We find  a good agreement between the measured  values of $g_1(t)$ and $g_2(t)$ at various crossover timescales and  the predictions of the reptation theory.


 Having determined the crossover time and length scales, we obtain the dynamic  exponents at various  scaling regimes.
 The best fits of $g_1(t)$ and $g_2(t)$ with power laws are shown by solid lines in Figs. \ref{fig17} and  \ref{fig18}. 
 In the first scaling regime, $\tau_0< t < \tau_e$, we find that $g_1(t)\approx  g_2(t) \propto t^{0.6}$ and the scaling exponent is larger than $1/2$. This difference  could again be attributed to the finite persistence length of chains as $N_e/\ell_K \approx  6$ is not large enough to fall in the random walk regime. In the second regime $\tau_e< t < \tau_R$, the obtained exponents of  $g_1(t)$ and $g_2(t)$ are 0.36 and 0.38 for $N=300$ and 0.31 and 0.35
 for $N=500$. These values   are noticeably larger than the  predicted $1/4$ exponent. These deviations are probably because chains are not long enough to fully develop this reptation regime. These results suggest that much longer chains  with  $N/N_e \gg 1$ are needed to observe the predicted exponent in the asymptotic limit \cite{Putz2000}. Nevertheless, upon increasing the chain length, the obtained exponents for the  second scaling regime slightly improve. In the third regime, $\tau_R< t < \tau_d$,    $g_1(t) \propto t^{0.55}$ and the exponents  of both chain lengths are slightly larger than $1/2$. In this regime,  $g_2(t) \propto t^{0.3}$ for $N=300$ and $g_2(t) \propto t^{0.32}$ for $N=500$.  Unlike the  theoretical predictions   $g_2(t)$ scaling significantly differs from that of $g_1(t)$.  This discrepancy is also most likely because  $N/N_e$ is not sufficiently large; the fit region includes the broad crossover to the asymptotic plateau.
  At the disengagement time,  we find that  the relation   $g_1(\tau_d)\approx g_3(\tau_d)$ roughly holds. For $N=300$, we can observe the free diffusion regime of
 $g_1(t)$  beyond $\tau_d$  where    $g_1(t)\approx g_3(t)$ and for  the $g_2(t>\tau_d)$, we observe the expected plateau value of $2 \langle R_g^2(N) \rangle$  for nearly two third of a decade.   For $N=500$, much longer simulation times are required to obtain reliable results for $t>\tau_d$.   
 
 Finally, we examine the scaling behavior of the mean square displacement of the center of mass of polymers. In contrast to the predictions of reptation 
 theory, we observe  two distinct scaling regimes for $t< \tau_R$. The first one corresponds to $ \tau_0< t < \tau_e$ 
 where $g_3(t)$ exhibits a subdiffusive behavior with  $g_3(t) \propto t^{0.76}$ for $N=300$ and  $g_3(t) \propto t^{0.78}$ for $N=500$.
 Similar to the short chains, this subdiffusive behavior results from the coupling to time-dependent viscoelatic relaxation of the melt
 background \cite{Farago1,Farago2}.   At timescales $\tau_e< t < \tau_R$, we observe a second scaling regime where  $g_3(t) \propto t^{0.68}$ for $N=300$ and  $g_3(t) \propto t^{0.62}$ for $N=500$. These exponents again are notably larger than the theoretical predictions due to the finite size of the chains, but as expected, they systematically decrease with chain length.

 \section*{Conclusions}
 
  We have investigated the static and dynamic  properties of  polymer melts of a  locally semiflexible bead-spring model known as the CG-PVA model \cite{Meyer2001,Meyer2002}. The main distinctive 
 feature of this model system is a triple-well  bending potential that leads to polymer crystallization upon cooling of the melt. 
 We have equilibrated polymer melts  with  chain lengths $5\le N \le 500$. 
  The results for the  long chains allow us to determine  the Kuhn length $\ell_K$,  the persistence length $\ell_p$ and the entanglement length $N_e$ of the model polymers accurately  as summarized in Table II.
  We note that the relation   $\ell_K\approx 2\ell_p$ holds  for the semiflexible CG-PVA polymers.   
   
   \begin{table}[h] 
\begin{tabular}{ | c | c | c | c |  c|}
 \hline
  &   &   &   &     \\
   
$\ell_b/\sigma$ & $\langle \cos \theta \rangle$ & $C_{\infty}$= $\ell_K/\ell_b$ & $\ell_p/\ell_b$ & $N_e$ \\
 &   &   &   &     \\
  \hline
   &   &   &   &     \\
  $0.497 $ & $0.699 $  & $5.70$   & $2.85 $& $36.5 $ \\
   $\pm 0.002$ & $ \pm 0.005$  & $\pm 0.04$   & $ \pm 0.01$& $ \pm 0.5$ \\
   &    &   &  &   \\
   \hline
   \end{tabular}
  \caption{Summary of  characteristic features of CG-PVA model   
where $ \ell_b $ is the average bond length, $\theta$ is the angle between two successive bonds, $\ell_p$, $\ell_K$ and $N_e$ are the  persistence length,   Kuhn length  and
entanglement length, respectively. }
 \end{table}

 We have also examined the validity of Flory's ideality hypothesis for this model
  system. Our detailed examination shows that the conformations of CG-PVA polymers agree with many of the theoretical predictions for ideal chains.
 Notably,  long polymer melts with $N>50$ follow the scaling relations $\langle R_e^2 \rangle/\langle R_g^2 \rangle=6$ and $\langle R_e^2 \rangle \propto N$  valid for ideal chains.
The probability distribution functions of the reduced end-to-end distance $r_e=(R_e^2/ \langle R_e^2 \rangle)^{1/2}$  and the reduced  gyration radius
$r_g=(R_g^2/ \langle R_g^2 \rangle)^{1/2}$    for chain lengths $N \ge 100$ also collapse on universal master curves that are well described by the theoretical distributions  for the 
ideal chains.  

Investigating the intrachain correlations, we find evidences for deviations from ideality.  However, these non-Gaussian corrections   are rather small and do not affect most of 
the large-scale conformational features. The mean square internal distances of short chains  show an excellent agreement with  the predictions of the generalized freely rotating chain model \cite{Flory} whereas those of longer chains are slightly larger.   
The observed   swelling of longer chains reflects an incomplete screening of excluded volume interactions and is  most likely related to the  correlation hole effect \cite{polymerDeGennes,Wittmer2007a}. We also compare the non-Gaussian parameter of CG-PVA polymers with that of the freely rotating chain model. The agreement is rather good and we only observe some small deviations for long
chains.

Carefully examining the bond-bond correlations of long chains at $T=1\approx 1.1T_c$, we do not observe any  long-range bond-bond correlations. 
Instead, the triple-well angle-bending potential leads to weak anti-correlations for  curvilinear distances of about $n \approx20 $.
The observed anti-correlation is a precursor of chain backfolding in the melt at temperatures slightly above the crystallization temperature. This feature is further enhanced  in the semicrystalline state  and it is  a signature of chain-folded structures \cite{Meyer2002,SaraLMC}.  The anti-correlations suppress  long-range bond-bond correlations but the long-range intrachain correlations are visible in the second Legendre polynomial of  the cosine of angles between the bonds, $\langle P_2[ \cos \theta(n)]\rangle $, that exhibit a power law decay in analogy to the reports for 2D polymer melts \cite{Meyer2010}.

 Moreover, we have investigated in detail the intrachain and interchain structure factors of different chain lengths. The interchain structure factor is almost independent of the chain length whereas the intrachain structure factor $S_c(q)$ depends
 on $N$ as expected.  We find that  $S_c(q)$ of sufficiently long  CG-PVA polymer are well-described by the Debye function for length scales larger than the Kuhn length. The agreement with the Debye function improves upon increase of $N$. Notably,
 we observe a plateau in the Kratky plot for the range $2< qR_g < 20$. Our  results  are in contrast with the findings for fully flexible chains that exhibit significant deviations from the Debye function at intermediate wave-vectors \cite{Wittmer2007a,
 formfactor2007,Hsu2014}. But, they agree with the recent findings that increasing the bending stiffness of the chains in a melt,  irrespective of details, improves the agreement with the 
 ideal-chain limit \cite{Kremer2016}.

 Using the primitive path analysis, we have  determined the average entanglement length $N_e$ of long and equilibrated chains  and we have compared  the original polymer paths with their primitive paths.
 Probing the bond-bond orientational correlation function and the mean square internal distance of primitive paths, we confirm the assumption that polymers behave nearly as Gaussian chains along their primitive paths.
 Notably, the Kuhn length  of the primitive path $\ell_K^{(pp)}$ is more than twice the $\ell_K$ of the original path.
 The average bond length of primitive paths follows a Gaussian distribution and the peak of the first moment of the entanglement length probability distribution  
 agrees with the  average entanglement length. 

 Investigating the segmental motion of entangled  polymer melts, we  observe several scaling regimes as predicted by the reptation theory. The crossover time and length scales between distinct scaling regimes agree with the predictions of the reptation theory  \cite{polymerDoi,KGmodel}. However the dynamical exponents of monomer mean square displacements are different from
 those of the theory most probably due to the finite persistence length and size of polymers.  The mean square displacement of center of mass of polymers
 also exhibits an anomalous diffusion at the timescale where the Rouse and reptation  theories predict a Fickian diffusion.  However, our results are qualitatively consistent
 with the more recent theories \cite{Farago1,Farago2} that  attribute the subdiffusive motion of polymers center of mass to viscoelastic hydrodynamic interactions. It still remains an open question, how a finite persistence length of polymers  affects quantitatively the dynamic exponents of center of mass particularly in the reptation regime.


 

\section*{acknowledgments}
   S.~J.-F.  is grateful to Jean-Louis Barrat  Hendrik Meyer, Kurt Kremer and Hsiao-Ping Hsu for insightful  discussions.
She also acknowledges financial support from the German Research Foundation (http://www.dfg.de) within SFB TRR 146 (http://trr146.de). 
The computations were performed using the Froggy platform of the CIMENT infrastructure supported by the Rhone-Alpes region (Grant No. CPER07-13 CIRA) and the Equip@Meso Project (Reference 337 No. ANR-10-EQPX-29-01) and  the supercomputer clusters Mogon I and II at Johannes Gutenberg University Mainz (hpc.uni-mainz.de).

\bibliography{melt.bib}

\begin{thebibliography}{49}%
\makeatletter
\providecommand \@ifxundefined [1]{%
 \@ifx{#1\undefined}
}%
\providecommand \@ifnum [1]{%
 \ifnum #1\expandafter \@firstoftwo
 \else \expandafter \@secondoftwo
 \fi
}%
\providecommand \@ifx [1]{%
 \ifx #1\expandafter \@firstoftwo
 \else \expandafter \@secondoftwo
 \fi
}%
\providecommand \natexlab [1]{#1}%
\providecommand \enquote  [1]{``#1''}%
\providecommand \bibnamefont  [1]{#1}%
\providecommand \bibfnamefont [1]{#1}%
\providecommand \citenamefont [1]{#1}%
\providecommand \href@noop [0]{\@secondoftwo}%
\providecommand \href [0]{\begingroup \@sanitize@url \@href}%
\providecommand \@href[1]{\@@startlink{#1}\@@href}%
\providecommand \@@href[1]{\endgroup#1\@@endlink}%
\providecommand \@sanitize@url [0]{\catcode `\\12\catcode `\$12\catcode
  `\&12\catcode `\#12\catcode `\^12\catcode `\_12\catcode `\%12\relax}%
\providecommand \@@startlink[1]{}%
\providecommand \@@endlink[0]{}%
\providecommand \url  [0]{\begingroup\@sanitize@url \@url }%
\providecommand \@url [1]{\endgroup\@href {#1}{\urlprefix }}%
\providecommand \urlprefix  [0]{URL }%
\providecommand \Eprint [0]{\href }%
\providecommand \doibase [0]{http://dx.doi.org/}%
\providecommand \selectlanguage [0]{\@gobble}%
\providecommand \bibinfo  [0]{\@secondoftwo}%
\providecommand \bibfield  [0]{\@secondoftwo}%
\providecommand \translation [1]{[#1]}%
\providecommand \BibitemOpen [0]{}%
\providecommand \bibitemStop [0]{}%
\providecommand \bibitemNoStop [0]{.\EOS\space}%
\providecommand \EOS [0]{\spacefactor3000\relax}%
\providecommand \BibitemShut  [1]{\csname bibitem#1\endcsname}%
\let\auto@bib@innerbib\@empty
\bibitem [{\citenamefont {Rubinstein}\ and\ \citenamefont
  {Colby}(2003)}]{Rubinstein}%
  \BibitemOpen
  \bibfield  {author} {\bibinfo {author} {\bibfnamefont {M.}~\bibnamefont
  {Rubinstein}}\ and\ \bibinfo {author} {\bibfnamefont {R.~H.}\ \bibnamefont
  {Colby}},\ }\href@noop {} {\emph {\bibinfo {title} {Polymer Physics}}}\
  (\bibinfo  {publisher} {Oxford University Press, Oxford},\ \bibinfo {year}
  {2003})\BibitemShut {NoStop}%
\bibitem [{\citenamefont {Mandelkern}(1990)}]{semicrys}%
  \BibitemOpen
  \bibfield  {author} {\bibinfo {author} {\bibfnamefont {L.}~\bibnamefont
  {Mandelkern}},\ }\href {\doibase 10.1021/ar00179a006} {\bibfield  {journal}
  {\bibinfo  {journal} {Accounts of Chemical Research}\ }\textbf {\bibinfo
  {volume} {23}},\ \bibinfo {pages} {380} (\bibinfo {year} {1990})},\ \Eprint
  {http://arxiv.org/abs/https://doi.org/10.1021/ar00179a006}
  {https://doi.org/10.1021/ar00179a006} \BibitemShut {NoStop}%
\bibitem [{\citenamefont {Keller}(1968)}]{Keller}%
  \BibitemOpen
  \bibfield  {author} {\bibinfo {author} {\bibfnamefont {A.}~\bibnamefont
  {Keller}},\ }\href@noop {} {\bibfield  {journal} {\bibinfo  {journal}
  {Reports on Progress in Physics}\ }\textbf {\bibinfo {volume} {31}},\
  \bibinfo {pages} {623} (\bibinfo {year} {1968})}\BibitemShut {NoStop}%
\bibitem [{\citenamefont {edited~by J.-U.~Sommer}\ and\ \citenamefont
  {Reiter}(2003)}]{Sommercrys}%
  \BibitemOpen
  \bibfield  {author} {\bibinfo {author} {\bibnamefont {edited~by
  J.-U.~Sommer}}\ and\ \bibinfo {author} {\bibfnamefont {G.}~\bibnamefont
  {Reiter}},\ }\href@noop {} {\bibfield  {journal} {\bibinfo  {journal}
  {Lecture Notes in Physics}\ }\textbf {\bibinfo {volume} {606}} (\bibinfo
  {year} {2003})}\BibitemShut {NoStop}%
\bibitem [{\citenamefont {Flory}(1956)}]{Florycrys}%
  \BibitemOpen
  \bibfield  {author} {\bibinfo {author} {\bibfnamefont {P.}~\bibnamefont
  {Flory}},\ }\href {\doibase 10.1098/rspa.1956.0015} {\bibfield  {journal}
  {\bibinfo  {journal} {Proceedings of the Royal Society of London A:
  Mathematical, Physical and Engineering Sciences}\ }\textbf {\bibinfo {volume}
  {234}},\ \bibinfo {pages} {60} (\bibinfo {year} {1956})}\BibitemShut
  {NoStop}%
\bibitem [{\citenamefont {Olmsted}\ \emph {et~al.}(1998)\citenamefont
  {Olmsted}, \citenamefont {Poon}, \citenamefont {McLeish}, \citenamefont
  {Terrill},\ and\ \citenamefont {Ryan}}]{Olmsted}%
  \BibitemOpen
  \bibfield  {author} {\bibinfo {author} {\bibfnamefont {P.~D.}\ \bibnamefont
  {Olmsted}}, \bibinfo {author} {\bibfnamefont {W.~C.~K.}\ \bibnamefont
  {Poon}}, \bibinfo {author} {\bibfnamefont {T.~C.~B.}\ \bibnamefont
  {McLeish}}, \bibinfo {author} {\bibfnamefont {N.~J.}\ \bibnamefont
  {Terrill}}, \ and\ \bibinfo {author} {\bibfnamefont {A.~J.}\ \bibnamefont
  {Ryan}},\ }\href {\doibase 10.1103/PhysRevLett.81.373} {\bibfield  {journal}
  {\bibinfo  {journal} {Phys. Rev. Lett.}\ }\textbf {\bibinfo {volume} {81}},\
  \bibinfo {pages} {373} (\bibinfo {year} {1998})}\BibitemShut {NoStop}%
\bibitem [{\citenamefont {Sushko}\ \emph {et~al.}(2001)\citenamefont {Sushko},
  \citenamefont {van~der Schoot},\ and\ \citenamefont {Michels}}]{DFTPaul}%
  \BibitemOpen
  \bibfield  {author} {\bibinfo {author} {\bibfnamefont {N.}~\bibnamefont
  {Sushko}}, \bibinfo {author} {\bibfnamefont {P.}~\bibnamefont {van~der
  Schoot}}, \ and\ \bibinfo {author} {\bibfnamefont {M.~A.~J.}\ \bibnamefont
  {Michels}},\ }\href {\doibase 10.1063/1.1404390} {\bibfield  {journal}
  {\bibinfo  {journal} {The Journal of Chemical Physics}\ }\textbf {\bibinfo
  {volume} {115}},\ \bibinfo {pages} {7744} (\bibinfo {year}
  {2001})}\BibitemShut {NoStop}%
\bibitem [{\citenamefont {McCoy}\ \emph {et~al.}(1991)\citenamefont {McCoy},
  \citenamefont {Honnell}, \citenamefont {Schweizer},\ and\ \citenamefont
  {Curro}}]{DFT1}%
  \BibitemOpen
  \bibfield  {author} {\bibinfo {author} {\bibfnamefont {J.~D.}\ \bibnamefont
  {McCoy}}, \bibinfo {author} {\bibfnamefont {K.~G.}\ \bibnamefont {Honnell}},
  \bibinfo {author} {\bibfnamefont {K.~S.}\ \bibnamefont {Schweizer}}, \ and\
  \bibinfo {author} {\bibfnamefont {J.~G.}\ \bibnamefont {Curro}},\ }\href
  {\doibase 10.1063/1.461163} {\bibfield  {journal} {\bibinfo  {journal} {The
  Journal of Chemical Physics}\ }\textbf {\bibinfo {volume} {95}},\ \bibinfo
  {pages} {9348} (\bibinfo {year} {1991})}\BibitemShut {NoStop}%
\bibitem [{\citenamefont {Oxtoby}(2002)}]{DFTreview}%
  \BibitemOpen
  \bibfield  {author} {\bibinfo {author} {\bibfnamefont {D.~W.}\ \bibnamefont
  {Oxtoby}},\ }\href@noop {} {\bibfield  {journal} {\bibinfo  {journal} {Annu.
  Rev. Mater. Res.}\ }\textbf {\bibinfo {volume} {32}},\ \bibinfo {pages} {39}
  (\bibinfo {year} {2002})}\BibitemShut {NoStop}%
\bibitem [{\citenamefont {Meyer}\ and\ \citenamefont
  {M\"uller-Plathe}(2001)}]{Meyer2001}%
  \BibitemOpen
  \bibfield  {author} {\bibinfo {author} {\bibfnamefont {H.}~\bibnamefont
  {Meyer}}\ and\ \bibinfo {author} {\bibfnamefont {F.}~\bibnamefont
  {M\"uller-Plathe}},\ }\href@noop {} {\bibfield  {journal} {\bibinfo
  {journal} {The Journal of Chemical Physics}\ }\textbf {\bibinfo {volume}
  {115}},\ \bibinfo {pages} {7807} (\bibinfo {year} {2001})}\BibitemShut
  {NoStop}%
\bibitem [{\citenamefont {Reith}\ \emph {et~al.}(2001)\citenamefont {Reith},
  \citenamefont {Meyer},\ and\ \citenamefont {M\"uller-Plathe}}]{Reith}%
  \BibitemOpen
  \bibfield  {author} {\bibinfo {author} {\bibfnamefont {D.}~\bibnamefont
  {Reith}}, \bibinfo {author} {\bibfnamefont {H.}~\bibnamefont {Meyer}}, \ and\
  \bibinfo {author} {\bibfnamefont {F.}~\bibnamefont {M\"uller-Plathe}},\
  }\href {\doibase 10.1021/ma001499k} {\bibfield  {journal} {\bibinfo
  {journal} {Macromolecules}\ }\textbf {\bibinfo {volume} {34}},\ \bibinfo
  {pages} {2335} (\bibinfo {year} {2001})},\ \Eprint
  {http://arxiv.org/abs/https://doi.org/10.1021/ma001499k}
  {https://doi.org/10.1021/ma001499k} \BibitemShut {NoStop}%
\bibitem [{\citenamefont {Jabbari-Farouji}\ \emph
  {et~al.}(2015{\natexlab{a}})\citenamefont {Jabbari-Farouji}, \citenamefont
  {Rottler}, \citenamefont {Lame}, \citenamefont {Makke}, \citenamefont
  {Perez},\ and\ \citenamefont {Barrat}}]{SaraMacro}%
  \BibitemOpen
  \bibfield  {author} {\bibinfo {author} {\bibfnamefont {S.}~\bibnamefont
  {Jabbari-Farouji}}, \bibinfo {author} {\bibfnamefont {J.}~\bibnamefont
  {Rottler}}, \bibinfo {author} {\bibfnamefont {O.}~\bibnamefont {Lame}},
  \bibinfo {author} {\bibfnamefont {A.}~\bibnamefont {Makke}}, \bibinfo
  {author} {\bibfnamefont {M.}~\bibnamefont {Perez}}, \ and\ \bibinfo {author}
  {\bibfnamefont {J.~L.}\ \bibnamefont {Barrat}},\ }\href@noop {} {\bibfield
  {journal} {\bibinfo  {journal} {ACS Macro Letters}\ }\textbf {\bibinfo
  {volume} {4}},\ \bibinfo {pages} {147} (\bibinfo {year}
  {2015}{\natexlab{a}})}\BibitemShut {NoStop}%
\bibitem [{\citenamefont {Jabbari-Farouji}\ \emph {et~al.}(2017)\citenamefont
  {Jabbari-Farouji}, \citenamefont {Lame}, \citenamefont {Perez}, \citenamefont
  {Rottler},\ and\ \citenamefont {Barrat}}]{SaraPRL2017}%
  \BibitemOpen
  \bibfield  {author} {\bibinfo {author} {\bibfnamefont {S.}~\bibnamefont
  {Jabbari-Farouji}}, \bibinfo {author} {\bibfnamefont {O.}~\bibnamefont
  {Lame}}, \bibinfo {author} {\bibfnamefont {M.}~\bibnamefont {Perez}},
  \bibinfo {author} {\bibfnamefont {J.}~\bibnamefont {Rottler}}, \ and\
  \bibinfo {author} {\bibfnamefont {J.-L.}\ \bibnamefont {Barrat}},\ }\href
  {\doibase 10.1103/PhysRevLett.118.217802} {\bibfield  {journal} {\bibinfo
  {journal} {Phys. Rev. Lett.}\ }\textbf {\bibinfo {volume} {118}},\ \bibinfo
  {pages} {217802} (\bibinfo {year} {2017})}\BibitemShut {NoStop}%
\bibitem [{\citenamefont {Triandafilidi}\ \emph {et~al.}(2016)\citenamefont
  {Triandafilidi}, \citenamefont {Rottler},\ and\ \citenamefont
  {Hatzikiriakos}}]{Joerg2016}%
  \BibitemOpen
  \bibfield  {author} {\bibinfo {author} {\bibfnamefont {V.}~\bibnamefont
  {Triandafilidi}}, \bibinfo {author} {\bibfnamefont {J.}~\bibnamefont
  {Rottler}}, \ and\ \bibinfo {author} {\bibfnamefont {S.~G.}\ \bibnamefont
  {Hatzikiriakos}},\ }\href@noop {} {\bibfield  {journal} {\bibinfo  {journal}
  {Journal of Polymer Science Part B: Polymer Physics}\ ,\ \bibinfo {pages}
  {2318}} (\bibinfo {year} {2016})}\BibitemShut {NoStop}%
\bibitem [{\citenamefont {Luo}\ and\ \citenamefont {Sommer}(2013)}]{Luo2013}%
  \BibitemOpen
  \bibfield  {author} {\bibinfo {author} {\bibfnamefont {C.}~\bibnamefont
  {Luo}}\ and\ \bibinfo {author} {\bibfnamefont {J.}~\bibnamefont {Sommer}},\
  }\href@noop {} {\bibfield  {journal} {\bibinfo  {journal} {ACS Macro
  Letters}\ }\textbf {\bibinfo {volume} {2}},\ \bibinfo {pages} {31} (\bibinfo
  {year} {2013})}\BibitemShut {NoStop}%
\bibitem [{\citenamefont {Luo}\ and\ \citenamefont {Sommer}(2016)}]{Luo2016}%
  \BibitemOpen
  \bibfield  {author} {\bibinfo {author} {\bibfnamefont {C.}~\bibnamefont
  {Luo}}\ and\ \bibinfo {author} {\bibfnamefont {J.-U.}\ \bibnamefont
  {Sommer}},\ }\href {\doibase 10.1021/acsmacrolett.5b00668} {\bibfield
  {journal} {\bibinfo  {journal} {ACS Macro Letters}\ }\textbf {\bibinfo
  {volume} {5}},\ \bibinfo {pages} {30} (\bibinfo {year} {2016})}\BibitemShut
  {NoStop}%
\bibitem [{\citenamefont {Vettorel}\ \emph {et~al.}(2007)\citenamefont
  {Vettorel}, \citenamefont {Meyer}, \citenamefont {Baschnagel},\ and\
  \citenamefont {Fuchs}}]{Vettorel2007}%
  \BibitemOpen
  \bibfield  {author} {\bibinfo {author} {\bibfnamefont {T.}~\bibnamefont
  {Vettorel}}, \bibinfo {author} {\bibfnamefont {H.}~\bibnamefont {Meyer}},
  \bibinfo {author} {\bibfnamefont {J.}~\bibnamefont {Baschnagel}}, \ and\
  \bibinfo {author} {\bibfnamefont {M.}~\bibnamefont {Fuchs}},\ }\href@noop {}
  {\bibfield  {journal} {\bibinfo  {journal} {Phys. Rev. E}\ }\textbf {\bibinfo
  {volume} {75}},\ \bibinfo {pages} {041801} (\bibinfo {year}
  {2007})}\BibitemShut {NoStop}%
\bibitem [{\citenamefont {Flory}(1969)}]{Flory}%
  \BibitemOpen
  \bibfield  {author} {\bibinfo {author} {\bibfnamefont {P.~J.}\ \bibnamefont
  {Flory}},\ }\href@noop {} {\emph {\bibinfo {title} {Statistical Mechanics of
  Chain Molecules}}}\ (\bibinfo  {publisher} {Wiley, New York},\ \bibinfo
  {year} {1969})\BibitemShut {NoStop}%
\bibitem [{\citenamefont {Doi}\ and\ \citenamefont
  {Edwards}(1986)}]{polymerDoi}%
  \BibitemOpen
  \bibfield  {author} {\bibinfo {author} {\bibfnamefont {M.}~\bibnamefont
  {Doi}}\ and\ \bibinfo {author} {\bibfnamefont {S.~F.}\ \bibnamefont
  {Edwards}},\ }\href@noop {} {\emph {\bibinfo {title} {The Theory of Polymer
  Dynamics}}}\ (\bibinfo  {publisher} {Clarendon Press, Oxford},\ \bibinfo
  {year} {1986})\BibitemShut {NoStop}%
\bibitem [{\citenamefont {Wittmer}\ \emph {et~al.}(2004)\citenamefont
  {Wittmer}, \citenamefont {Meyer}, \citenamefont {Baschnageland},
  \citenamefont {Johner}, \citenamefont {Obukhov}, \citenamefont {Mattioni},
  \citenamefont {Muller},\ and\ \citenamefont {Semenov}}]{Wittmer2004}%
  \BibitemOpen
  \bibfield  {author} {\bibinfo {author} {\bibfnamefont {J.~P.}\ \bibnamefont
  {Wittmer}}, \bibinfo {author} {\bibfnamefont {H.}~\bibnamefont {Meyer}},
  \bibinfo {author} {\bibfnamefont {J.}~\bibnamefont {Baschnageland}}, \bibinfo
  {author} {\bibfnamefont {A.}~\bibnamefont {Johner}}, \bibinfo {author}
  {\bibfnamefont {S.}~\bibnamefont {Obukhov}}, \bibinfo {author} {\bibfnamefont
  {L.}~\bibnamefont {Mattioni}}, \bibinfo {author} {\bibfnamefont
  {M.}~\bibnamefont {Muller}}, \ and\ \bibinfo {author} {\bibfnamefont {A.~N.}\
  \bibnamefont {Semenov}},\ }\href@noop {} {\bibfield  {journal} {\bibinfo
  {journal} {Physical Review Letters}\ }\textbf {\bibinfo {volume} {93}},\
  \bibinfo {pages} {147801} (\bibinfo {year} {2004})}\BibitemShut {NoStop}%
\bibitem [{\citenamefont {Wittmer}\ \emph
  {et~al.}(2007{\natexlab{a}})\citenamefont {Wittmer}, \citenamefont
  {Beckrich}, \citenamefont {Johner}, \citenamefont {Semenov}, \citenamefont
  {Obukhov}, \citenamefont {Meyer},\ and\ \citenamefont
  {Baschnagel}}]{Wittmer2007a}%
  \BibitemOpen
  \bibfield  {author} {\bibinfo {author} {\bibfnamefont {J.~P.}\ \bibnamefont
  {Wittmer}}, \bibinfo {author} {\bibfnamefont {P.}~\bibnamefont {Beckrich}},
  \bibinfo {author} {\bibfnamefont {A.}~\bibnamefont {Johner}}, \bibinfo
  {author} {\bibfnamefont {A.~N.}\ \bibnamefont {Semenov}}, \bibinfo {author}
  {\bibfnamefont {S.}~\bibnamefont {Obukhov}}, \bibinfo {author} {\bibfnamefont
  {H.}~\bibnamefont {Meyer}}, \ and\ \bibinfo {author} {\bibfnamefont
  {J.}~\bibnamefont {Baschnagel}},\ }\href@noop {} {\bibfield  {journal}
  {\bibinfo  {journal} {Europhysics Letters}\ }\textbf {\bibinfo {volume}
  {77}},\ \bibinfo {pages} {56003} (\bibinfo {year}
  {2007}{\natexlab{a}})}\BibitemShut {NoStop}%
\bibitem [{\citenamefont {Wittmer}\ \emph
  {et~al.}(2007{\natexlab{b}})\citenamefont {Wittmer}, \citenamefont
  {Beckrich}, \citenamefont {Meyer}, \citenamefont {Cavallo}, \citenamefont
  {Johner},\ and\ \citenamefont {Baschnagel}}]{Wittmer2007b}%
  \BibitemOpen
  \bibfield  {author} {\bibinfo {author} {\bibfnamefont {J.~P.}\ \bibnamefont
  {Wittmer}}, \bibinfo {author} {\bibfnamefont {P.}~\bibnamefont {Beckrich}},
  \bibinfo {author} {\bibfnamefont {H.}~\bibnamefont {Meyer}}, \bibinfo
  {author} {\bibfnamefont {A.}~\bibnamefont {Cavallo}}, \bibinfo {author}
  {\bibfnamefont {A.}~\bibnamefont {Johner}}, \ and\ \bibinfo {author}
  {\bibfnamefont {J.}~\bibnamefont {Baschnagel}},\ }\href@noop {} {\bibfield
  {journal} {\bibinfo  {journal} {Physical Review E}\ }\textbf {\bibinfo
  {volume} {76}},\ \bibinfo {pages} {011803} (\bibinfo {year}
  {2007}{\natexlab{b}})}\BibitemShut {NoStop}%
\bibitem [{\citenamefont {Beckrich}\ \emph {et~al.}(2007)\citenamefont
  {Beckrich}, \citenamefont {Johner}, \citenamefont {Semenov}, \citenamefont
  {Obukhov}, \citenamefont {Benoit},\ and\ \citenamefont
  {Wittmer}}]{formfactor2007}%
  \BibitemOpen
  \bibfield  {author} {\bibinfo {author} {\bibfnamefont {P.}~\bibnamefont
  {Beckrich}}, \bibinfo {author} {\bibfnamefont {A.}~\bibnamefont {Johner}},
  \bibinfo {author} {\bibfnamefont {A.~N.}\ \bibnamefont {Semenov}}, \bibinfo
  {author} {\bibfnamefont {S.~P.}\ \bibnamefont {Obukhov}}, \bibinfo {author}
  {\bibfnamefont {H.~C.}\ \bibnamefont {Benoit}}, \ and\ \bibinfo {author}
  {\bibfnamefont {J.~P.}\ \bibnamefont {Wittmer}},\ }\href@noop {} {\bibfield
  {journal} {\bibinfo  {journal} {Macromolecules}\ }\textbf {\bibinfo {volume}
  {40}},\ \bibinfo {pages} {3805} (\bibinfo {year} {2007})}\BibitemShut
  {NoStop}%
\bibitem [{\citenamefont {Hsu}(2014)}]{Hsu2014}%
  \BibitemOpen
  \bibfield  {author} {\bibinfo {author} {\bibfnamefont {H.-P.}\ \bibnamefont
  {Hsu}},\ }\href@noop {} {\bibfield  {journal} {\bibinfo  {journal} {J. Chem.
  Phys.}\ }\textbf {\bibinfo {volume} {141}},\ \bibinfo {pages} {164903}
  (\bibinfo {year} {2014})}\BibitemShut {NoStop}%
\bibitem [{\citenamefont {Semenov}(2010)}]{Semenov}%
  \BibitemOpen
  \bibfield  {author} {\bibinfo {author} {\bibfnamefont {A.~N.}\ \bibnamefont
  {Semenov}},\ }\href {\doibase 10.1021/ma101465z} {\bibfield  {journal}
  {\bibinfo  {journal} {Macromolecules}\ }\textbf {\bibinfo {volume} {43}},\
  \bibinfo {pages} {9139} (\bibinfo {year} {2010})}\BibitemShut {NoStop}%
\bibitem [{\citenamefont {Hsu}\ and\ \citenamefont
  {Kremer}(2016)}]{Kremer2016}%
  \BibitemOpen
  \bibfield  {author} {\bibinfo {author} {\bibfnamefont {H.-P.}\ \bibnamefont
  {Hsu}}\ and\ \bibinfo {author} {\bibfnamefont {K.}~\bibnamefont {Kremer}},\
  }\href@noop {} {\bibfield  {journal} {\bibinfo  {journal} {J. Chem. Phys.}\
  }\textbf {\bibinfo {volume} {144}},\ \bibinfo {pages} {154907} (\bibinfo
  {year} {2016})}\BibitemShut {NoStop}%
\bibitem [{\citenamefont {de~Gennes}(1979)}]{polymerDeGennes}%
  \BibitemOpen
  \bibfield  {author} {\bibinfo {author} {\bibfnamefont {P.~G.}\ \bibnamefont
  {de~Gennes}},\ }\href@noop {} {\emph {\bibinfo {title} {Scaling Concepts in
  Polymer Physics}}}\ (\bibinfo  {publisher} {Cornell University Press,
  Itharca, New York},\ \bibinfo {year} {1979})\BibitemShut {NoStop}%
\bibitem [{\citenamefont {Meyer}\ and\ \citenamefont
  {M\"uller-Plathe}(2002)}]{Meyer2002}%
  \BibitemOpen
  \bibfield  {author} {\bibinfo {author} {\bibfnamefont {H.}~\bibnamefont
  {Meyer}}\ and\ \bibinfo {author} {\bibfnamefont {F.}~\bibnamefont
  {M\"uller-Plathe}},\ }\href@noop {} {\bibfield  {journal} {\bibinfo
  {journal} {Macromolecules}\ }\textbf {\bibinfo {volume} {35}},\ \bibinfo
  {pages} {1241} (\bibinfo {year} {2002})}\BibitemShut {NoStop}%
\bibitem [{\citenamefont {Jabbari-Farouji}\ \emph
  {et~al.}(2015{\natexlab{b}})\citenamefont {Jabbari-Farouji}, \citenamefont
  {Rottler}, \citenamefont {Lame}, \citenamefont {Makke}, \citenamefont
  {Perez},\ and\ \citenamefont {Barrat}}]{SaraLMC}%
  \BibitemOpen
  \bibfield  {author} {\bibinfo {author} {\bibfnamefont {S.}~\bibnamefont
  {Jabbari-Farouji}}, \bibinfo {author} {\bibfnamefont {J.}~\bibnamefont
  {Rottler}}, \bibinfo {author} {\bibfnamefont {O.}~\bibnamefont {Lame}},
  \bibinfo {author} {\bibfnamefont {A.}~\bibnamefont {Makke}}, \bibinfo
  {author} {\bibfnamefont {M.}~\bibnamefont {Perez}}, \ and\ \bibinfo {author}
  {\bibfnamefont {J.~L.}\ \bibnamefont {Barrat}},\ }\href@noop {} {\bibfield
  {journal} {\bibinfo  {journal} {Journal of Physics: Condensed Matter}\
  }\textbf {\bibinfo {volume} {27}},\ \bibinfo {pages} {194131} (\bibinfo
  {year} {2015}{\natexlab{b}})}\BibitemShut {NoStop}%
\bibitem [{\citenamefont {Plimpton}(1995)}]{LAMMPS}%
  \BibitemOpen
  \bibfield  {author} {\bibinfo {author} {\bibfnamefont {S.}~\bibnamefont
  {Plimpton}},\ }\href@noop {} {\bibfield  {journal} {\bibinfo  {journal}
  {Journal of Computational Physics}\ }\textbf {\bibinfo {volume} {117}},\
  \bibinfo {pages} {1} (\bibinfo {year} {1995})}\BibitemShut {NoStop}%
\bibitem [{\citenamefont {Honnell}\ \emph
  {et~al.}(1990{\natexlab{a}})\citenamefont {Honnell}, \citenamefont {Curro},\
  and\ \citenamefont {Schweizer}}]{FRC}%
  \BibitemOpen
  \bibfield  {author} {\bibinfo {author} {\bibfnamefont {K.~G.}\ \bibnamefont
  {Honnell}}, \bibinfo {author} {\bibfnamefont {J.~G.}\ \bibnamefont {Curro}},
  \ and\ \bibinfo {author} {\bibfnamefont {K.~S.}\ \bibnamefont {Schweizer}},\
  }\href@noop {} {\bibfield  {journal} {\bibinfo  {journal} {Macromolecules}\
  }\textbf {\bibinfo {volume} {23}},\ \bibinfo {pages} {3496} (\bibinfo {year}
  {1990}{\natexlab{a}})}\BibitemShut {NoStop}%
\bibitem [{\citenamefont {Honnell}\ \emph
  {et~al.}(1990{\natexlab{b}})\citenamefont {Honnell}, \citenamefont {Curro},\
  and\ \citenamefont {Schweizer}}]{Honnell}%
  \BibitemOpen
  \bibfield  {author} {\bibinfo {author} {\bibfnamefont {K.~G.}\ \bibnamefont
  {Honnell}}, \bibinfo {author} {\bibfnamefont {J.~G.}\ \bibnamefont {Curro}},
  \ and\ \bibinfo {author} {\bibfnamefont {K.~S.}\ \bibnamefont {Schweizer}},\
  }\href@noop {} {\bibfield  {journal} {\bibinfo  {journal} {Macromolecules.}\
  }\textbf {\bibinfo {volume} {23}},\ \bibinfo {pages} {3496} (\bibinfo {year}
  {1990}{\natexlab{b}})}\BibitemShut {NoStop}%
\bibitem [{\citenamefont {Porod}(1953)}]{Porod1953}%
  \BibitemOpen
  \bibfield  {author} {\bibinfo {author} {\bibfnamefont {G.}~\bibnamefont
  {Porod}},\ }\href {\doibase 10.1002/pol.1953.120100203} {\bibfield  {journal}
  {\bibinfo  {journal} {Journal of Polymer Science}\ }\textbf {\bibinfo
  {volume} {10}},\ \bibinfo {pages} {157} (\bibinfo {year} {1953})}\BibitemShut
  {NoStop}%
\bibitem [{\citenamefont {Meyer}\ \emph {et~al.}(2010)\citenamefont {Meyer},
  \citenamefont {Wittmer}, \citenamefont {Kreer}, \citenamefont {Johner},\ and\
  \citenamefont {Baschnagel}}]{Meyer2010}%
  \BibitemOpen
  \bibfield  {author} {\bibinfo {author} {\bibfnamefont {H.}~\bibnamefont
  {Meyer}}, \bibinfo {author} {\bibfnamefont {J.~P.}\ \bibnamefont {Wittmer}},
  \bibinfo {author} {\bibfnamefont {T.}~\bibnamefont {Kreer}}, \bibinfo
  {author} {\bibfnamefont {A.}~\bibnamefont {Johner}}, \ and\ \bibinfo {author}
  {\bibfnamefont {J.}~\bibnamefont {Baschnagel}},\ }\href {\doibase
  10.1063/1.3429350} {\bibfield  {journal} {\bibinfo  {journal} {The Journal of
  Chemical Physics}\ }\textbf {\bibinfo {volume} {132}},\ \bibinfo {pages}
  {184904} (\bibinfo {year} {2010})}\BibitemShut {NoStop}%
\bibitem [{\citenamefont {Schmidt}\ and\ \citenamefont
  {Stockmayer}(1984)}]{Koyama0}%
  \BibitemOpen
  \bibfield  {author} {\bibinfo {author} {\bibfnamefont {M.}~\bibnamefont
  {Schmidt}}\ and\ \bibinfo {author} {\bibfnamefont {W.}~\bibnamefont
  {Stockmayer}},\ }\href@noop {} {\bibfield  {journal} {\bibinfo  {journal}
  {Macromolecules}\ }\textbf {\bibinfo {volume} {17}},\ \bibinfo {pages} {509}
  (\bibinfo {year} {1984})}\BibitemShut {NoStop}%
\bibitem [{\citenamefont {Mansfield}(1986)}]{Koyama}%
  \BibitemOpen
  \bibfield  {author} {\bibinfo {author} {\bibfnamefont {M.~L.}\ \bibnamefont
  {Mansfield}},\ }\href@noop {} {\bibfield  {journal} {\bibinfo  {journal}
  {Macromolecules}\ }\textbf {\bibinfo {volume} {19}},\ \bibinfo {pages} {854}
  (\bibinfo {year} {1986})}\BibitemShut {NoStop}%
\bibitem [{\citenamefont {Fujita}\ and\ \citenamefont
  {Norisuye}(1970)}]{pdfRg}%
  \BibitemOpen
  \bibfield  {author} {\bibinfo {author} {\bibfnamefont {H.}~\bibnamefont
  {Fujita}}\ and\ \bibinfo {author} {\bibfnamefont {T.}~\bibnamefont
  {Norisuye}},\ }\href@noop {} {\bibfield  {journal} {\bibinfo  {journal} {J.
  Chem. Phys.}\ }\textbf {\bibinfo {volume} {52}},\ \bibinfo {pages} {1115}
  (\bibinfo {year} {1970})}\BibitemShut {NoStop}%
\bibitem [{\citenamefont {Lhuillier}(1988)}]{Lhuillier}%
  \BibitemOpen
  \bibfield  {author} {\bibinfo {author} {\bibfnamefont {D.}~\bibnamefont
  {Lhuillier}},\ }\href@noop {} {\bibfield  {journal} {\bibinfo  {journal} {J.
  Phys. France}\ }\textbf {\bibinfo {volume} {49}},\ \bibinfo {pages} {705}
  (\bibinfo {year} {1988})}\BibitemShut {NoStop}%
\bibitem [{\citenamefont {Vettorel}\ \emph {et~al.}(2010)\citenamefont
  {Vettorel}, \citenamefont {Besold},\ and\ \citenamefont {Kremer}}]{pdfRg1}%
  \BibitemOpen
  \bibfield  {author} {\bibinfo {author} {\bibfnamefont {T.}~\bibnamefont
  {Vettorel}}, \bibinfo {author} {\bibfnamefont {G.}~\bibnamefont {Besold}}, \
  and\ \bibinfo {author} {\bibfnamefont {K.}~\bibnamefont {Kremer}},\
  }\href@noop {} {\bibfield  {journal} {\bibinfo  {journal} {Soft Matter}\
  }\textbf {\bibinfo {volume} {6}},\ \bibinfo {pages} {2282} (\bibinfo {year}
  {2010})}\BibitemShut {NoStop}%
\bibitem [{\citenamefont {Hansen}\ and\ \citenamefont
  {McDonald}(1986)}]{Hansen}%
  \BibitemOpen
  \bibfield  {author} {\bibinfo {author} {\bibfnamefont {J.~P.}\ \bibnamefont
  {Hansen}}\ and\ \bibinfo {author} {\bibfnamefont {I.~R.}\ \bibnamefont
  {McDonald}},\ }\href@noop {} {\emph {\bibinfo {title} {Theory of Simple
  Liquids}}}\ (\bibinfo  {publisher} {Academic Press, London},\ \bibinfo {year}
  {1986})\BibitemShut {NoStop}%
\bibitem [{\citenamefont {Edwards}(1967)}]{Edwardstube}%
  \BibitemOpen
  \bibfield  {author} {\bibinfo {author} {\bibfnamefont {S.~F.}\ \bibnamefont
  {Edwards}},\ }\href@noop {} {\bibfield  {journal} {\bibinfo  {journal} {Proc.
  Phys. Soc.}\ }\textbf {\bibinfo {volume} {91}},\ \bibinfo {pages} {513}
  (\bibinfo {year} {1967})}\BibitemShut {NoStop}%
\bibitem [{\citenamefont {Kr\"oger}(2005)}]{Kroger}%
  \BibitemOpen
  \bibfield  {author} {\bibinfo {author} {\bibfnamefont {M.}~\bibnamefont
  {Kr\"oger}},\ }\href@noop {} {\bibfield  {journal} {\bibinfo  {journal}
  {Comput. Phys. Commun.}\ }\textbf {\bibinfo {volume} {168}},\ \bibinfo
  {pages} {209} (\bibinfo {year} {2005})}\BibitemShut {NoStop}%
\bibitem [{\citenamefont {R.~Hoy}(2009)}]{Hoy2009}%
  \BibitemOpen
  \bibfield  {author} {\bibinfo {author} {\bibfnamefont {M.~K.}\ \bibnamefont
  {R.~Hoy}, \bibfnamefont {K.~Foteinopoulou}},\ }\href@noop {} {\bibfield
  {journal} {\bibinfo  {journal} {Phys. Rev. E}\ }\textbf {\bibinfo {volume}
  {80}},\ \bibinfo {pages} {031803} (\bibinfo {year} {2009})}\BibitemShut
  {NoStop}%
\bibitem [{\citenamefont {Everaers}\ \emph {et~al.}(2004)\citenamefont
  {Everaers}, \citenamefont {Sukumaran}, \citenamefont {Grest}, \citenamefont
  {Svaneborg}, \citenamefont {Sivasubramanian},\ and\ \citenamefont
  {Kremer}}]{PPAEveraers}%
  \BibitemOpen
  \bibfield  {author} {\bibinfo {author} {\bibfnamefont {R.}~\bibnamefont
  {Everaers}}, \bibinfo {author} {\bibfnamefont {S.~K.}\ \bibnamefont
  {Sukumaran}}, \bibinfo {author} {\bibfnamefont {G.~S.}\ \bibnamefont
  {Grest}}, \bibinfo {author} {\bibfnamefont {C.}~\bibnamefont {Svaneborg}},
  \bibinfo {author} {\bibfnamefont {A.}~\bibnamefont {Sivasubramanian}}, \ and\
  \bibinfo {author} {\bibfnamefont {K.}~\bibnamefont {Kremer}},\ }\href
  {\doibase 10.1126/science.1091215} {\bibfield  {journal} {\bibinfo  {journal}
  {Science}\ }\textbf {\bibinfo {volume} {303}},\ \bibinfo {pages} {823}
  (\bibinfo {year} {2004})}\BibitemShut {NoStop}%
\bibitem [{\citenamefont {Barkema}\ \emph {et~al.}(2014)\citenamefont
  {Barkema}, \citenamefont {Panja},\ and\ \citenamefont {van
  Leeuwen}}]{Barkema2014}%
  \BibitemOpen
  \bibfield  {author} {\bibinfo {author} {\bibfnamefont {G.~T.}\ \bibnamefont
  {Barkema}}, \bibinfo {author} {\bibfnamefont {D.}~\bibnamefont {Panja}}, \
  and\ \bibinfo {author} {\bibfnamefont {J.~M.~J.}\ \bibnamefont {van
  Leeuwen}},\ }\href {http://stacks.iop.org/1742-5468/2014/i=11/a=P11008}
  {\bibfield  {journal} {\bibinfo  {journal} {Journal of Statistical Mechanics:
  Theory and Experiment}\ }\textbf {\bibinfo {volume} {2014}},\ \bibinfo
  {pages} {P11008} (\bibinfo {year} {2014})}\BibitemShut {NoStop}%
\bibitem [{\citenamefont {Farago}\ \emph {et~al.}(2012)\citenamefont {Farago},
  \citenamefont {Meyer}, \citenamefont {Baschnagel},\ and\ \citenamefont
  {Semenov}}]{Farago1}%
  \BibitemOpen
  \bibfield  {author} {\bibinfo {author} {\bibfnamefont {J.}~\bibnamefont
  {Farago}}, \bibinfo {author} {\bibfnamefont {H.}~\bibnamefont {Meyer}},
  \bibinfo {author} {\bibfnamefont {J.}~\bibnamefont {Baschnagel}}, \ and\
  \bibinfo {author} {\bibfnamefont {A.~N.}\ \bibnamefont {Semenov}},\ }\href
  {\doibase 10.1103/PhysRevE.85.051807} {\bibfield  {journal} {\bibinfo
  {journal} {Phys. Rev. E}\ }\textbf {\bibinfo {volume} {85}},\ \bibinfo
  {pages} {051807} (\bibinfo {year} {2012})}\BibitemShut {NoStop}%
\bibitem [{\citenamefont {Farago}\ \emph {et~al.}(2011)\citenamefont {Farago},
  \citenamefont {Meyer},\ and\ \citenamefont {Semenov}}]{Farago2}%
  \BibitemOpen
  \bibfield  {author} {\bibinfo {author} {\bibfnamefont {J.}~\bibnamefont
  {Farago}}, \bibinfo {author} {\bibfnamefont {H.}~\bibnamefont {Meyer}}, \
  and\ \bibinfo {author} {\bibfnamefont {A.~N.}\ \bibnamefont {Semenov}},\
  }\href {\doibase 10.1103/PhysRevLett.107.178301} {\bibfield  {journal}
  {\bibinfo  {journal} {Phys. Rev. Lett.}\ }\textbf {\bibinfo {volume} {107}},\
  \bibinfo {pages} {178301} (\bibinfo {year} {2011})}\BibitemShut {NoStop}%
\bibitem [{\citenamefont {Kremer}\ and\ \citenamefont {Grest}(1990)}]{KGmodel}%
  \BibitemOpen
  \bibfield  {author} {\bibinfo {author} {\bibfnamefont {K.}~\bibnamefont
  {Kremer}}\ and\ \bibinfo {author} {\bibfnamefont {G.~S.}\ \bibnamefont
  {Grest}},\ }\href@noop {} {\bibfield  {journal} {\bibinfo  {journal} {J.
  Chem. Phys.}\ }\textbf {\bibinfo {volume} {92}},\ \bibinfo {pages} {5057}
  (\bibinfo {year} {1990})}\BibitemShut {NoStop}%
\bibitem [{\citenamefont {P\"utz}\ \emph {et~al.}(2000)\citenamefont {P\"utz},
  \citenamefont {Kremer},\ and\ \citenamefont {Grest}}]{Putz2000}%
  \BibitemOpen
  \bibfield  {author} {\bibinfo {author} {\bibfnamefont {M.}~\bibnamefont
  {P\"utz}}, \bibinfo {author} {\bibfnamefont {K.}~\bibnamefont {Kremer}}, \
  and\ \bibinfo {author} {\bibfnamefont {G.~S.}\ \bibnamefont {Grest}},\ }\href
  {http://stacks.iop.org/0295-5075/49/i=6/a=735} {\bibfield  {journal}
  {\bibinfo  {journal} {EPL (Europhysics Letters)}\ }\textbf {\bibinfo {volume}
  {49}},\ \bibinfo {pages} {735} (\bibinfo {year} {2000})}\BibitemShut
  {NoStop}%
\end{thebibliography}%

\end{document}


@article{example,
  author={},
  title={},
  journal={},
  volume={},
  number={},
  pages={},
  url={},
  year={},
  abstract={}
}